\documentclass[a4paper,11pt]{article}
\pdfoutput=1 % if your are submitting a pdflatex (i.e. if you have
\usepackage{jheppub} % for details on the use of the package, please
\usepackage[T1]{fontenc} % if needed
\usepackage{amsmath}
\usepackage{amssymb}
\usepackage{slashed}
\usepackage{color}
\usepackage{float}
\usepackage{afterpage}
\usepackage{caption}
\usepackage{subcaption}
\usepackage[english]{babel}
\usepackage{comment}
\usepackage{tikz-feynman}
\tikzfeynmanset{compat=1.1.0}

\DeclareRobustCommand{\App}[1]{App.~\ref{#1}}
\DeclareRobustCommand{\Tab}[1]{Table~\ref{#1}}

\DeclareRobustCommand{\Fig}[1]{Fig.~\ref{#1}}
\DeclareRobustCommand{\Eq}[1]{Eq.~(\ref{#1})}
\DeclareRobustCommand{\Eqs}[2]{Eqs.~(\ref{#1}) and (\ref{#2})}

\definecolor{red1}{cmyk}{0,1,1,0.3}

\newcommand{\beq}{\begin{equation}}
\newcommand{\eeq}{\end{equation}}

\newcommand{\eg}{\textit{e.g.}}
\newcommand{\ie}{\textit{i.e.}}

\newcommand{\NPfield}{\phi}

\preprint{UCI-TR-2024-24}

\title{
Flavor at FASER: Discovering Light Scalars Beyond Minimal Flavor Violation
}
\author[a,b]{Reuven Balkin,}
\author[a]{Noam Burger,}
\author[c]{Jonathan L.~Feng,}
\author[a]{Yael Shadmi}
\affiliation[a]{Physics Department, Technion -- Israel Institute of Technology, Haifa 3200003, Israel}
\affiliation[b]{Department of Physics, University of California Santa Cruz and Santa Cruz Institute for Particle Physics, 1156 High St., Santa Cruz, CA 95064, USA}
\affiliation[c]{Department of Physics and Astronomy, University of California, Irvine, CA  92697 USA}
\emailAdd{rebalkin@ucsc.edu}
\emailAdd{noam.burger@campus.technion.ac.il}
\emailAdd{jlf@uci.edu}
\emailAdd{yshadmi@physics.technion.ac.il}
\abstract{
We study a simple class of \emph{flavored scalar} models, in which the couplings of a new light scalar to standard-model fermions are controlled by the flavor symmetry responsible for fermion masses and mixings. 
The scalar couplings are then aligned with the Yukawa matrices, with small but nonzero flavor-violating entries.
$D$-meson decays are an important source of scalar production in these models, in contrast to models assuming minimal flavor violation, in which $B$ and $K$ decays dominate. 
We show that FASER2 can probe large portions of the parameter space of the models, with  comparable numbers of scalars from $B$ and $D$ decays in some regions. 
If discovered, these particles will not only provide evidence of new physics, but they may also shed new light on the standard model flavor puzzle. 
Finally, the richness of theoretical models underscores the importance of model-independent interpretations.  
We therefore analyze the sensitivity of FASER and other experimental searches in terms of physical parameters:~(i) the branching fractions of  heavy mesons to the scalar, and (ii) $\tau/m$, where $\tau$ and $m$ are the scalar's lifetime and mass, respectively. 
The results are largely independent of the new particle's spin and can be used to extract constraints on a wide variety of models.
}
\begin{document} 

\maketitle
\flushbottom

%%%%%%%%%%%%%%%%%%%%%%%%%%%%%%%%%%%%%%%%%%%%%%%%%%%%%%%%%%%%%%%%
\section{Introduction}
\label{sec:intro-new}
%%%%%%%%%%%%%%%%%%%%%%%%%%%%%%%%%%%%%%%%%%%%%%%%%%%%%%%%%%%%%%%%

With little definitive guidance from theory, the masses and couplings of hypothetical sub-GeV particles can span many orders of magnitude. 
Still, there are two robust features of these particles that are dictated by experiment. 
First, their couplings to standard model (SM) particles must be small, implying long lifetimes in the absence of hidden-sector decays; and second, of these couplings, flavor-violating couplings are especially constrained.
The flavor constraints can be easily alleviated with the assumption of minimal flavor violation (MFV), namely, that the SM Yukawa couplings are the only source of flavor dependence.  

It is important to bear in mind, however, that the hierarchical patterns of quark and lepton masses and mixings remain one of the most puzzling aspects of the SM,  hinting at an underlying theory of flavor. 
Such a theory should give rise to the required non-trivial textures of the SM Yukawas.  
It therefore naturally also generates non-trivial textures in the couplings of new long-lived particles (LLPs) to fermions.  
The experimental signatures of such \emph{flavored LLPs}, as we will refer to them in the following, can be quite distinct from MFV predictions.  
In MFV models, the physical LLP-fermion coupling matrices---defined in the fermion mass basis---are diagonal.  
Flavor-violating couplings are then only generated by $W$-boson loops and are $G_F$-suppressed. 
Furthermore, since the dominant flavor-diagonal coupling is to the top quark, the loop induced flavor-violating couplings to up-type quarks are subleading, and LLPs from rare meson decays are mainly produced from $B$- or $K$-mesons. 
On the other hand, in flavored models, tree-level, flavor-violating LLP couplings to up quarks are typically present, and $D$-meson decays may provide an important source of $\NPfield$ production, in addition to $B$ or $K$ decays. 

Here we consider a simple example, featuring a \emph{flavored scalar} LLP~$\NPfield$ with the most general $\NPfield$ couplings to the SM, parametrized by a dimension-five effective theory (EFT) Lagrangian. 
In addition, we assume that the SM flavor structure arises from a $U(1)$ Froggatt-Nielsen symmetry~\cite{Froggatt:1978nt,Leurer:1992wg,Leurer:1993gy} broken by a spurion $\lambda\sim0.2$.  
For an appropriate choice of fermion $U(1)$ charges, the different entries of the Yukawa matrices are suppressed by different powers of $\lambda$, and the required hierarchies of the fermion masses and mixings are generated. 
The entries of the $\NPfield$-fermion coupling matrices also feature powers of $\lambda$, with the power completely determined by the $\NPfield$'s $U(1)$ charge. 
Taking the simplest choice, with $\NPfield$ neutral under the flavor symmetry, the $\NPfield$ couplings are aligned with the Yukawas---they are approximately diagonal when the Yukawas are diagonalized~\cite{Nir:1993mx,Leurer:1993gy}.
However, this alignment is only approximate, because, although the flavor symmetry determines the powers of $\lambda$, it does not determine their ${\cal O}(1)$ prefactors.
The flavor-violating processes may therefore be brought under control, as in Refs.~\cite{Nir:1993mx,Leurer:1993gy,Shadmi:2011hs}, but the flavor off-diagonal $\NPfield$ couplings, although small, are nonzero.

We will explore the discovery potential for flavored scalars at FASER~\cite{Feng:2017uoz,FASER:2018bac,FASER:2022hcn} and its planned upgrade, FASER2~\cite{Anchordoqui:2021ghd,Feng:2022inv,Adhikary:2024nlv}.  
These LHC experiments allow for essentially background-free observations of LLPs decaying into any visible final state, including hadrons, charged leptons, and photons.  
The discovery prospects for light scalars at FASER and FASER2 have been explored in several studies; see, \eg, Refs.~\cite{Feng:2017vli,Winkler:2018qyg,FASER:2018eoc,Boiarska:2019jym,Boiarska:2019vid,Okada:2019opp,Popa:2021fgy,Li:2021rzt,Araki:2022xqp,Kling:2022uzy,Ferber:2023iso}.  
These previous studies, and indeed, the effort of most large community studies in this area~\cite{Battaglieri:2017aum,Alimena:2019zri,Beacham:2019nyx}, have focused primarily on dark Higgs bosons and MFV couplings.    
The main assumption in these models is that the $\NPfield$ couplings to the SM arise from the $\phi$-Higgs mixing, and, in particular, that non-renormalizable $\phi$ couplings to SM fermions are subdominant (see \App{app:higgs_mixing}). 
This assumption highly constrains the production and decay of light scalars.  
There are studies that have considered other coupling patterns, \eg, to illustrate some simple alternatives or to explain an experimental anomaly~\cite{Batell:2017kty,Batell:2018fqo,Batell:2021xsi,Egana-Ugrinovic:2019wzj,Kling:2020mch}.
In this work, we show that interesting alternatives to MFV can also emerge as a consequence of symmetries in flavor theories motivated by the observed masses and mixings of the SM fermions.  
This work highlights the importance of signals that are not dominant in the MFV framework, and also implies that the discovery of new scalars, in addition to providing evidence for BSM physics, may also shed light on the SM flavor problem, one of the most fundamental problems in particle physics today.

The importance of considering non-MFV couplings in QCD axion and axion-like particle (ALP) searches, and the relative paucity of constraints from the up-sector was stressed, \eg, in Ref.~\cite{MartinCamalich:2020dfe}. 
The general flavor structure of ALP models and the resulting phenomenology was studied in many examples, including Refs.~\cite{Kamenik:2011vy,Bauer:2020jbp,Carmona:2021seb,Bauer:2021mvw,MartinCamalich:2020dfe,Greljo:2024evt,Goudzovski:2022vbt}.
Indeed, ALPs naturally arise as the axions of broken flavor symmetries~\cite{Wilczek:1982rv,Feng:1997tn}.
In particular, ALPs from
broken Froggatt-Nielsen symmetries were studied in Refs.~\cite{MartinCamalich:2020dfe,Carmona:2021seb,Greljo:2024evt}.
Our philosophy in this paper is in some sense orthogonal. 
We do not assume that the LLP is related in any way to the flavor theory. 
Rather, our main observation is that if there is some LLP and some flavor theory, the latter automatically governs the flavor structure of LLP couplings.

We study two variants of the $\NPfield$ EFT Lagrangian. 
In the first, which we refer to as FN (Froggatt-Nielsen), the $\NPfield$ particle couples to all fermion species, while in the other, which we refer to as FNU (Froggatt-Nielsen--up), the $\NPfield$ field only couples to up-quarks at dimension 5.
These two variants can arise in different UV models.
Since the flavor symmetry only determines the spurion dependence of the $\NPfield$ couplings and not their ${\cal{O}}(1)$ prefactors, there are large variations in the experimental  signatures of the flavored scalar models.
In some regions of the model parameter space, $B$ and $D$ decays provide comparable sources of $\NPfield$ particles.
If the $\NPfield$ particles are produced in sufficient numbers, detailed studies could disentangle their different production and decay modes, providing welcome new insights on SM flavor.

As our analysis demonstrates, LLP models and signatures are significantly richer than the commonly considered MFV framework. 
This underscores the importance of a model-independent interpretation of experimental searches. 
Indeed, constraints on LLPs have been derived in terms of physical parameters, namely the branching fractions for LLP production and the LLP lifetime $\tau$~\cite{Dobrich:2018jyi}. 
However, the relevant quantity for a given experiment is the lifetime in the laboratory frame$~p_0\tau/m$, where $p_0$ is the characteristic momentum of LLPs reaching the detector, and $m$ is the LLP's mass.  
Therefore, as we show, the sensitivity of different experiments has ``flat directions'' corresponding to the combination $\tau/m$. 
Plotting the reach in terms of $\tau/m$ removes these flat directions, and efficiently encodes the constraints on a wide variety of models.
In many experiments, including FASER, where the LLP is highly boosted, the constraints are essentially independent of the LLP spin.
Our model-independent results, therefore, can easily be used to determine bounds and discovery reaches for LLPs in many specific models.

This paper is organized as follows.  The model-independent analysis of the FASER sensitivity and existing constraints  appears in Section~\ref{sec:modindep}. 
In Section~\ref{sec:models}, we present the flavored scalar models under consideration, focusing on a specific Froggatt-Nielsen model.   
The experimental signatures of flavored LLPs from $B$- and $D$-meson decays at FASER are discussed in Section~\ref{sec:pheno}, where we also show existing constraints.  
We conclude in Section~\ref{sec:conclusions}.  
\App{app:decays} details the equations we use for the different decay channels.  
\App{app:higgs_mixing} addresses the sizes of $\phi$-Higgs mixing and dimension-5 contributions in flavored models and in dark Higgs models.

%%%%%%%%%%%%%%%%%%%%%%%%%%%%%%%%%%%%%%%%%%%%
%%%%%%%%%%%%%%%%%%%%%%%%%%%%%%%%%%%%%%%%%%%%%%
\section{Model-Independent Reach}
\label{sec:modindep}
%%%%%%%%%%%%%%%%%%%%%%%%%%%%%%%%%%%%%%%%%%%%
%%%%%%%%%%%%%%%%%%%%%%%%%%%%%%%%%%%%%%%%%%%%%%

Consider a new light particle $\NPfield$ produced in a two-body decay, 
\begin{align}
    M \; \to \; M' \,\NPfield \,,
    \label{eq:M_decay}
\end{align}
where $M$ and $M^\prime$ are SM mesons.
This particle can be detected at a forward-placed detector like FASER if it decays within the designated decay volume to visible final states. 
When estimating the reach of a given experiment, it is useful to quantify it in terms of the physical properties of $\NPfield$, namely its mass $m_\NPfield$, lifetime $\tau_\NPfield$, the branching fractions $\text{Br}(M\to M'\,\NPfield)$, and 
\begin{align}
\sum_{i\in \text{visible}}\text{Br}(\NPfield\to i) \equiv \text{Br}(\NPfield\to \text{visible})\,,
\end{align}
where the visible final states are determined by the experiment. 
In this section, we advance a model-independent approach that, given various simplifying assumptions, allows for simple recasts of sensitivity reaches or current bounds onto the parameter space of any SM extension, as advocated in Refs.~\cite{LHCb:2015nkv,LHCb:2016awg,Dobrich:2018jyi}.
Note that in this work, we assume for simplicity that $\NPfield$ is a scalar. However, the spin does not play an important role in the following discussion, as long as the $M$ particles are not produced polarized, as is usually the case. 
Then, even if $M$ has a spin, calculating its spin-averaged decay rate would wash away any information regarding the spin of $\NPfield$. 

Let us start our discussion by disentangling the different dependencies and identifying the properties of $\NPfield$ that most affect the experimental sensitivity. 
The number of signal events at a given experiment is given schematically by
\begin{align}
    N_\NPfield \sim N_M  \cdot \text{Br}(M\to M'\,\NPfield) \cdot \mathcal{A}(m_\NPfield) \cdot P_{\text{decay}}(\tau_\NPfield/m_\NPfield)\sum_{i\in \text{visible}} \text{Br}(\NPfield\to i)\cdot \epsilon_i\,.
\end{align}
$N_M$ is the number of $M$ and $\overline{M}$ mesons produced, which depends on the experiment; for $pp$ collisions, it depends on the center-of-mass energy and luminosity, and for fixed target experiments, it depends on the number of protons on target.  
The geometrical acceptance factor $\mathcal{A}(m_\NPfield)$ is the fraction of $\NPfield$ particles that are produced with trajectories that intersect the detector, which depends strongly on the angular coverage of the detector.  
It depends weakly on $m_\NPfield$, through its effect on the decay kinematics of $M$, which becomes negligible in the limit $m_\NPfield \ll m_M$.\footnote{Throughout this work we make the simplifying assumption that the trajectory of $\NPfield$ is a good proxy for the trajectory of its decay products, or, in other words, that if $\NPfield$ decays inside the decay volume, then its decay products are sufficiently collimated that they pass through the detector.  For larger masses, taking the opening angle of the decay products into account would typically produce at most an $\mathcal{O}(10\%)$ reduction in signal, which we therefore neglect. 
On the other hand, a very light and therefore very boosted $\NPfield$ would lead to highly collimated decay products, which could escape detection, depending on the experiment selection criteria.}

The probability that a $\NPfield$ passing through the detector decays in the decay volume is given by
\begin{equation}
    P_{\text{decay}}(\tau_\NPfield/m_\NPfield) \approx \exp(-L/d_{\text{lab}})\, \left[ 1- \exp(-\Delta/d_{\text{lab}}) \right] \,,
    \label{eq:P_decay}
\end{equation}
where $L$ is the distance between the interaction point (where $\NPfield$ is produced) and the start of the decay volume, and $\Delta$ is the length of the decay volume.  
The distance traveled by $\NPfield$ in the lab frame is
\begin{equation}
    d_{\text{lab}}= \langle p_\NPfield \rangle \frac{c\tau_\NPfield}{m_\NPfield} \,,
    \label{eq:d_Lab}
\end{equation}
where $\langle p_\NPfield \rangle$ is the typical momentum of a $\NPfield$ particle in the acceptance region. 
Due to energy-momentum conservation, $\NPfield$ roughly inherits the kinematic properties of $M$ in the lab frame up to small $m_\NPfield$-dependent corrections. 
Therefore, $\langle p_\NPfield \rangle$ strongly depends on the spectrum of $M$, which in turn depends on the particular experimental setup.  

When the mesons are produced at rest, one naturally finds that $\left< p_{\NPfield} \right> \lesssim m_M$. 
The scaling with the heavy meson mass may also persist for production at high center-of-mass energies $\sqrt{s} \gg m_M$, such as at the LHC. 
In this case, the production spectrum of heavy mesons peaks at a transverse momentum $p_M^T \equiv p_M \sin \theta \sim m_M$~\cite{Feng:2017vli}, where $\theta$ is the scattering angle of the meson.
Thus, the average momentum satisfies $\left< p_{\NPfield} \right> \approx m_M / \theta$ for $\theta \gtrsim m_M/\sqrt{s}$,  which applies for the acceptance angle in FASER2.
For $\theta \lesssim m_M/\sqrt{s}$, which applies for the acceptance angles in FASER, the typical  
$\left<p_\NPfield\right>$ can be evaluated numerically.

\Eqs{eq:P_decay}{eq:d_Lab} imply that the sensitivity has a strong exponential dependence on the value of $c\tau_\NPfield/m_\NPfield$.  They also imply that for a given experiment, peak sensitivity is reached for
\begin{align}
\frac{c\tau_\NPfield}{m_\NPfield} \sim  \frac{L}{\langle p_\NPfield \rangle}\,,
\label{eq:peak_sens}
\end{align}
where the right-hand side is fixed for a given experiment, and is the most relevant combination of experimental parameters for many LLP searches~\cite{Feng:2017uoz}.
The discussion above is applicable to other similar experimental setups searching for long-lived particles, such as CHARM~\cite{CHARM:1985anb} and SHiP~\cite{Aberle:2839677}.

Lastly, the signal yield is proportional to the sum over all the experimentally visible states $\{i\}$, weighted by the relevant branching fraction $\text{Br}(\NPfield\to i)$ and the experimental efficiency $\epsilon_i$. We fix $\epsilon_i = 1$ throughout this work. 

 As an illustration of this model-independent method, we can estimate the strength of existing bounds from CHARM and the projected reach for FASER and FASER2 in terms of physical parameters for $\{M,M'\}=\{B,K\}$ and, for the first time, $\{M,M'\}=\{D,\pi\}$. 
 The proposed SHiP experiment can potentially probe large parts of the parameter space under consideration if one assumes zero background events.
 A more reliable estimate of the sensitivity of SHiP requires a study of the background rates, which is beyond the scope of this work.
 The geometries assumed in the calculation for the various experiments are listed in \Tab{tab:geometries}, and the total $B$- and $D$-meson yields for the various experiments are listed in \Tab{tab:B_D_yields}.
\begin{table}[bp]
  \centering
    \begin{tabular}{| c| c| c | c| c |c|}
    \hline
    Experiment & $L$ (m) & $\Delta$ (m)& $S\,(\text{m}^2)$ & $\delta$ (m) &  Reference
    \\
    \hline \hline
    FASER & 480 & 1.5 & $\pi\,(0.1)^2$ & -  & \cite{FASER:2022hcn}
    \\
    \hline
    FASER2 & 650 & 10 & $\pi\, (1)^2$ & -  & \cite{FASER2}
    \\
    \hline
    CHARM & 480  & 35 & $3\times 3$ & 3.5 &  \cite{CHARM:1985anb}
    \\
    \hline
    \end{tabular}
  \caption{Geometries of the various experiments considered in this work. $L$ denotes the distance between the interaction point and the beginning of the decay volume, $\Delta$ is the length of the decay volume, $S$ is the transverse area of the detector.  FASER and FASER2 are centered on the beam collision axis, and for CHARM, $\delta$ is the distance between the bottom of the decay volume and the beam collision axis.}
  \label{tab:geometries}
\end{table}

\begin{table}[tbph]
  \centering
    \begin{tabular}{| c| c| c | c| c |c|c|}
    \hline
    Experiment & $\sigma(pN\to\bar{b}b)$ (pb) & $\sigma(pN\to\bar{c}c)$ (pb) & $\mathcal{L}$  (pb$^{-1}$) & $N_B$ & $N_D$
    \\
    \hline \hline
    FASER &  $ 2.5\times 10^{8}$ & $ 3.8\times 10^{9}$ & $3\times10^{5}$ & $1.5\times 10^{14}$  & $2.2\times 10^{15}$  
    \\
    \hline
    FASER2 &  $ 2.5\times 10^{8}$ & $ 3.8\times 10^{9}$ & $3\times10^{6}$ & $1.5\times 10^{15}$  & $2.2\times 10^{16}$  
    \\
    \hline
    CHARM  & $1.8\times 10^{5}$  & $1.1\times10^{9}$ &  $2.2\times 10^{6}$ &  $7.9\times10^{11}$ & $5.1\times 10^{15}$
    \\
    \hline
    \end{tabular}
\caption{Total production cross sections, luminosities, and total yields used in this work, where $N=p$ for the production cross section at the LHC for FASER and FASER2, and $N=\text{Cu}$ for the production cross section for CHARM. The listed LHC cross sections are for a single hemisphere, and the CHARM production cross sections and effective luminosity are determined as described in the text.  }
  \label{tab:B_D_yields}
\end{table}

For FASER and FASER2 at the LHC, the $B$- and $D$-meson spectra and total production cross sections from $pp$ collisions were calculated using FONLL~\cite{Cacciari:1998it,Cacciari:2001td}, and we neglect secondary production~\cite{Jodlowski:2019ycu}. 
The cross sections at the LHC are consistent with simulation results found in the literature; see, \eg, Ref.~\cite{Carrer:2003ac}.  
The total yield is $N_B = 2 \, \sigma(pp \to \bar{b}b) \, {\cal L}$, with a similar formula for $N_D$, where the factor of two accounts for the two heavy mesons produced in every $p p \to \bar{b} b$ interaction, and the cross section is the cross section for one hemisphere, as noted in \Tab{tab:B_D_yields}. 

For the fixed target experiment CHARM, $N_M = 2\, \sigma(pN\to \bar{q}q)\, \mathcal{L}_{\text{\tiny eff}}$.  We determine the cross section by starting with the FONLL results for $\sigma(pp\to \bar{q}q)$ at the appropriate center-of-mass energy, and then scaling them under the assumption $\sigma(p\,N \to \bar{q}{q}) = A\, \sigma(p\,p \to \bar{q}{q})$~\cite{Lourenco:2006vw}, where $A=63$ for copper, which was the target material in CHARM. 
We again neglect secondary production~\cite{CERN-SHiP-NOTE-2015-009}, which typically corrects the production for thick targets by an $\mathcal{O}(1)$ factor, and leads to somewhat different values of $\langle p_\NPfield \rangle$.  
The effective luminosity for fixed-target experiments is given by $\mathcal{L}_{\text{\tiny eff}} = N_{\text{\tiny POT}}/\sigma(p\,N\to\text{anything})$, where $N_{\text{\tiny POT}}= 2.4\times10^{18}$ for CHARM, and we take $\sigma(pN\to\text{anything})= 53\,\text{mb}\,A^{0.77}$~\cite{Carvalho:2003pza}. 
The resulting fixed target production rate for a $400\,$GeV proton beam is consistent with observations~\cite{Lourenco:2006vw}.  

\begin{figure}[tbp]
    \centering    \includegraphics[width=0.48\textwidth]{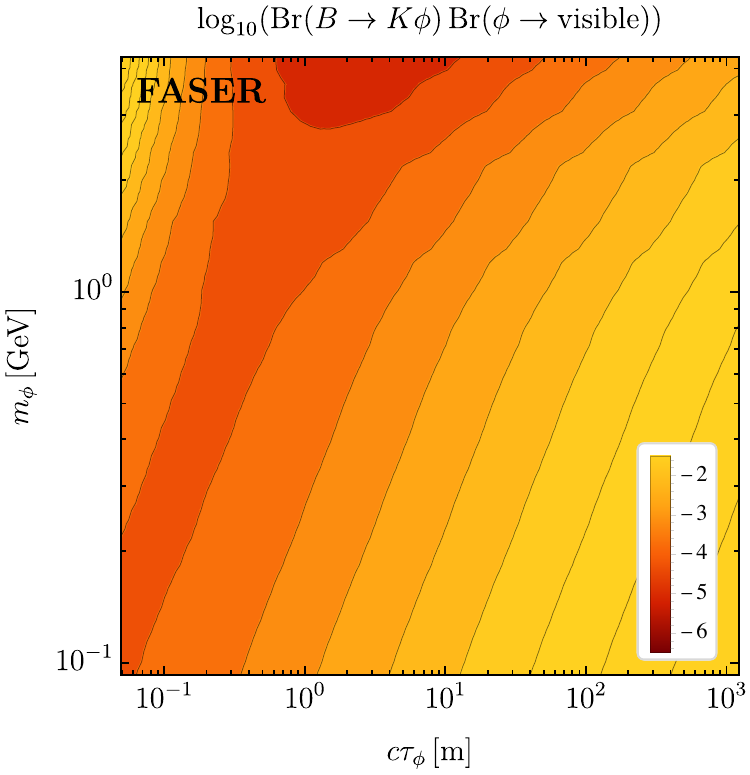} 
    \hfill
    \includegraphics[width=0.48\textwidth]{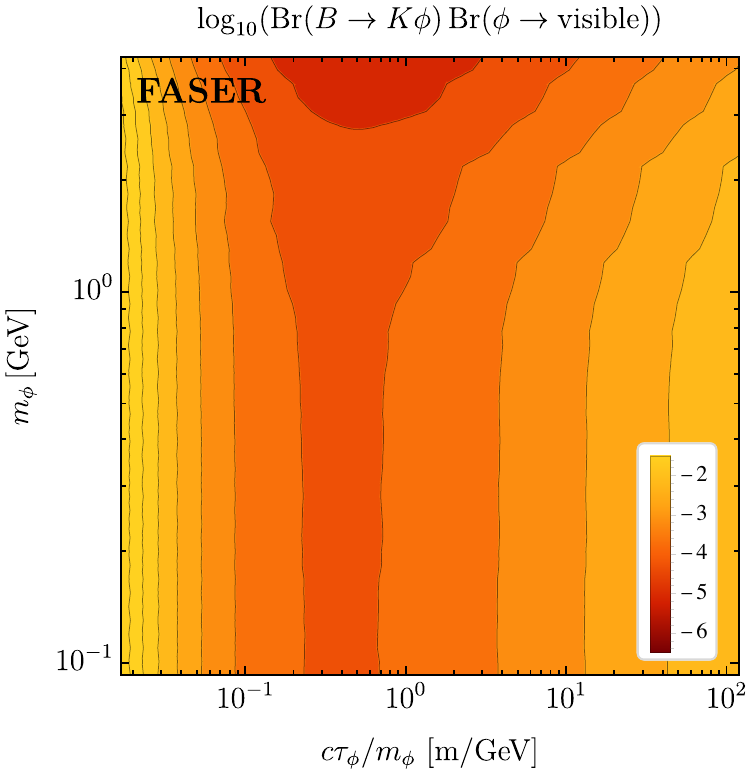}
\caption{Left: Projected sensitivity of FASER to $\text{Br}(B\to K\,\NPfield)\cdot\text{Br}(\NPfield\to \text{visible})$ in the $(m_\NPfield,\tau_\NPfield)$ plane. Right: Same, but in the $(m_\NPfield, c\tau_\NPfield/m_\NPfield)$ plane.}
    \label{fig:FASER_B_meson_flat_direction}
\end{figure}

The model-independent sensitivity to $\text{Br}(M\to M'\,\NPfield)\cdot\text{Br}(\NPfield\to \text{visible})$ can be expressed as a function of $(m_\NPfield,\tau_\NPfield)$. 
Following the discussion above, we note that by changing variables to $(m_\NPfield, c\tau_\NPfield / m_\NPfield)$, one finds only a weak dependence on the value of $m_\NPfield$, namely, an effectively flat direction in the parameter space.
To illustrate this, we plot in \Fig{fig:FASER_B_meson_flat_direction} our projected sensitivity of FASER to $\text{Br}(B\to K\,\NPfield)\cdot\text{Br}(\NPfield\to \text{visible})$ in the $(m_\NPfield,\tau_\NPfield)$ and $(m_\NPfield,c\tau_\NPfield / m_\NPfield)$ planes in the left and right panels, respectively. 
For $m_\NPfield \ll m_B$, we find that the sensitivity only depends on $c\tau_\NPfield/m_\NPfield$. 
As $m_\NPfield$ increases and approaches the kinematic threshold $m_B-m_K$, the decay products of the $B$ meson become increasingly collinear with the direction of the $B$, and a larger fraction of $\NPfield$ particle are contained in the acceptance region, which leads to increased sensitivity.
However, in this region the branching fraction $\text{Br}(B\to K\,\NPfield)$ starts getting suppressed due to the increasingly small volume of available phase space.  
As a consistency check, let us verify that the peak sensitivity is indeed around the prediction of \Eq{eq:peak_sens}. 
One finds that in the acceptance region of FASER, the typical momentum of a $B$ meson is $\langle p_B \rangle_{\text{\tiny FASER}} \sim \langle p_\NPfield \rangle_{\text{\tiny FASER}} \sim 600\,\text{GeV}$. 
With $L_{\text{\tiny FASER}}=480\,\text{m}$, we find that $L_{\text{\tiny FASER}}/\langle p_\NPfield \rangle_{\text{\tiny FASER}}\sim 0.8\,$m/GeV, consistent up to $\mathcal{O}(1)$ factors with our results, which peak at $c \tau_\NPfield/{m_\NPfield} \sim 0.4\,$m/GeV. 

%%%%%%%%%%%
%%%%%%%%%%%

%%%%%%%%%%%%%%%
\begin{figure}[tbp]
\centering    
\includegraphics[width=0.8\textwidth]{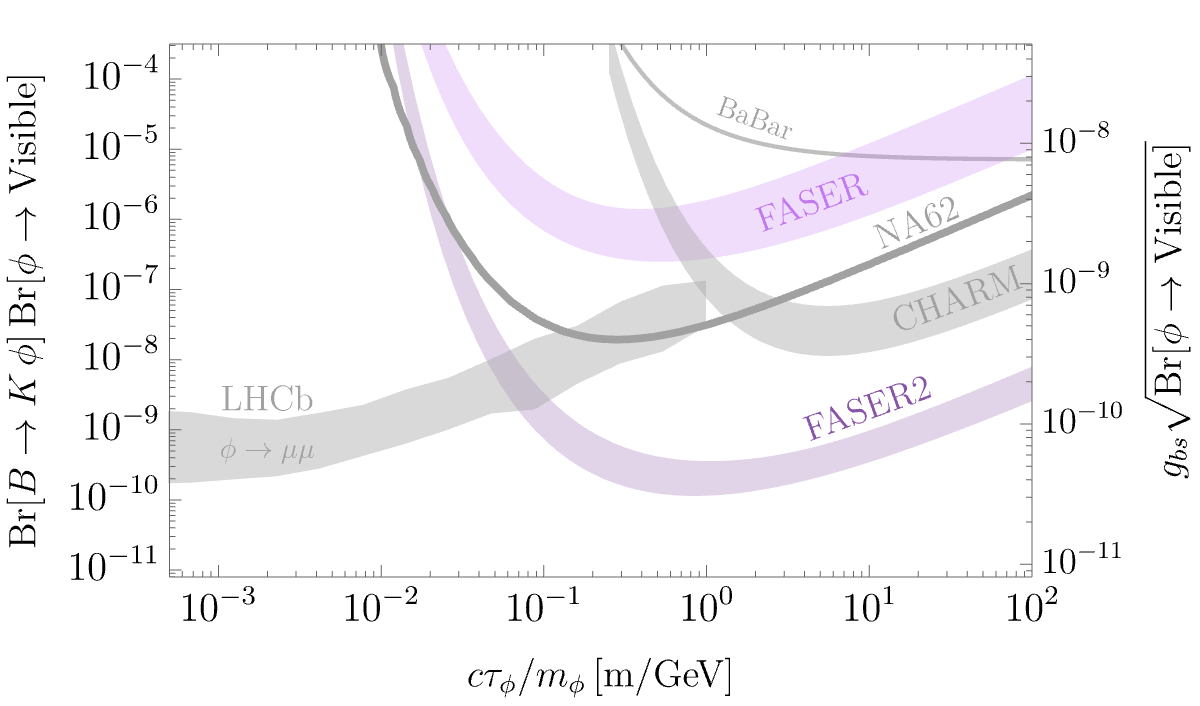} 
\caption{Existing bounds (gray) and projected sensitivities (purple) for $\text{Br}(B\to K\,\NPfield)\cdot\text{Br}(\NPfield\to \text{visible})$ as a function of $c\tau_\NPfield/m_{\NPfield}$.  The LHCb bounds has no sensitivity around SM resonance masses, but this is obscured in the $c\tau_\NPfield/m_{\NPfield}$ plot.  The $y$ axis on the right shows the effective coupling defined in~\Eq{eq:effcoupling}. The widths of the bands correspond to varying $m_\NPfield$; see text for more details. } 
\label{fig:B_model_independent_reach}
\end{figure}
 %%%%%%%%%%%%%%%

%%%%%%%%%%
\begin{figure}[tbp]
\centering 
\includegraphics[width=0.8\textwidth]{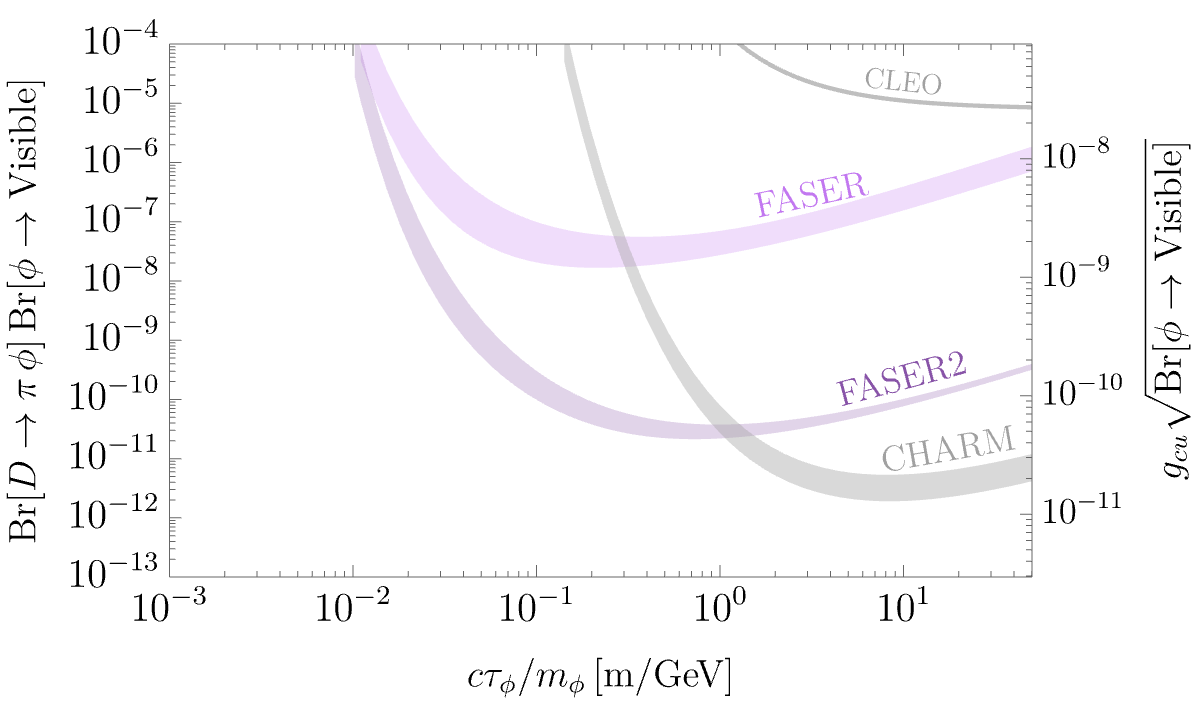} 
\caption{Existing bounds (gray) and projected sensitivities (purple) for $\text{Br}(D\to \pi\,\NPfield)\cdot\text{Br}(\NPfield\to \text{visible})$ as a function of $c\tau_\NPfield/m_{\NPfield}$. The $y$ axis on the right shows the effective coupling defined in~\Eq{eq:effcoupling}. The widths of the bands correspond to varying $m_\NPfield$; see text for more details.}
\label{fig:D_model_independent_reach}
\end{figure}
%%%%%%%%%%%%

The model-independent sensitivities for $B$ and $D$ decays as a function of $c\tau_\NPfield/m_\NPfield$ are plotted in \Fig{fig:B_model_independent_reach} and \Fig{fig:D_model_independent_reach}, respectively.
The projected sensitivity lines are defined by requiring 3 signal events, assuming zero background events.
In \Fig{fig:B_model_independent_reach}, we plot existing bounds from the search $B\to K\,\NPfield( \mu^+ \mu^-)$ performed by LHCb~\cite{LHCb:2015nkv,LHCb:2016awg} and  NA62~\cite{NA62:2023qyn}, as well as the recast of the BaBar search~\cite{BaBar:2013npw,MartinCamalich:2020dfe} for missing energy, generalized to finite lifetime,
\begin{align}
    \text{Br}[B \to K\NPfield] < (7.1\times 10^{-6})P^{-1}\;\;\text{  with  }\;\; P \equiv \exp \left[\frac{-L_{\text{\tiny BaBar}}/\langle p_\NPfield \rangle_{\text{\tiny BaBar}}}{c\tau_\NPfield/m_\NPfield} \right]\,.  \label{eq:BaBar_search}   
\end{align}
Since the BaBar search was for missing energy signatures, we weight this bound with the fraction of particles that decay outside the BaBar detector.
Note that we assume for simplicity that $\NPfield$ does not decay invisibly;
to include that possibility, one can further generalize the bound by taking $P \to P+\text{Br}[\NPfield \to \text{invis.}](1-P)$.
The $B$ mesons produced in BaBar were non-relativistic, and so we estimate $\langle p_\NPfield \rangle_{\text{\tiny BaBar}} \approx m_B/2$ and take $L_{\text{\tiny BaBar}} \approx 3\,$m.
Similarly, in \Fig{fig:D_model_independent_reach}, we plot the recast of the CLEO bound  $ 8\times 10^{-6}$~\cite{CLEO:2008ffk,MartinCamalich:2020dfe}, weighted by a similar exponential factor with $\langle p_\NPfield \rangle_{\text{\tiny CLEO}} \approx m_D/2$ and $L_{\text{\tiny CLEO}}\approx 3\,$m.

The bands in \Fig{fig:B_model_independent_reach} and \Fig{fig:D_model_independent_reach} 
are obtained by varying $m_\NPfield$ from $\sim m_{M}-m_{M'}$ (bottom of band) to $m_\NPfield \ll m_{M}$ (top of band).
The different experimental sensitivities peak at different values
of $c \tau_\NPfield / m_\NPfield$, depending on the value of $L/\langle p_\NPfield \rangle$ for a given experiment. Thus, although $L$ is comparable
for CHARM and FASER, the latter probes shorter lifetimes because the $\NPfield$ particles are typically produced with a larger boost. 
In fact, as explained above, the peak sensitivity in each experiment slightly shifts between \Fig{fig:B_model_independent_reach} and \Fig{fig:D_model_independent_reach} because of the different momentum distributions
of $\NPfield$'s from $B$ and $D$ decays. 
This feature can be exploited to disentangle the two production channels, as we discuss in Section~\ref{sec:pheno}.
For short lifetimes $d_{\text{lab}} \ll L$, the sensitivities are exponentially suppressed, while for $d_{\text{lab}} \gg L$ the decay probability $P_{\text{decay}}(\tau_\NPfield/m_\NPfield) \approx \Delta/d_{\text{lab}}$, and the sensitivity declines as $(\tau_\NPfield/m_\NPfield)^{-1}$.
As expected, LHCb exhibits a qualitatively different behavior at short lifetimes; once the $\NPfield$ particle decays at length scales below the vertex resolution and is prompt at the scale of the detector, the sensitivity loses its dependence on the value of $\tau_\NPfield$.
The NA62 sensitivity in \Fig{fig:B_model_independent_reach} is based on four similar masses given in Ref.~\cite{NA62:2023qyn}, and therefore the thin band does not capture the additional mass dependence.  
Lastly, since the BaBar and CLEO are missing energy searches, the mass dependence comes in only in the exponential decay factor.

It is instructive to display the sensitivity in terms of the effective off-diagonal vector-like coupling $g_{bs}$ ($g_{cu}$) responsible for the $B$ ($D$) decays, where we use the parton-level estimate,
\begin{align}\label{eq:effcoupling}
    \text{Br}(M \to M' \phi) \approx \frac{g^2 m_{M}}{32 \pi\Gamma_{M}}\,,
\end{align}
assuming $m_M \gg m_\phi,m_{M'}$. We show the effective coupling on the $y$-axis on the right-hand side of the plots.

This model-independent analysis has yielded some very interesting conclusions.  
From \Fig{fig:B_model_independent_reach}, we see that for scalars produced in $B$ decays, FASER cannot probe new parameter space that is not already excluded by NA62. 
On the other hand, FASER2, with its $\sim 10^3$ times larger decay volume and 10 times larger luminosity, is roughly a factor of $10^4$ more sensitive, and probes new parameter space for $c \tau_\phi / m_\phi \gtrsim 0.1$.  
Compared to CHARM, FASER2 benefits from a much larger center-of-mass energy, and so $B$ production is not kinematically suppressed.  
For scalars produced in $D$ decays, \Fig{fig:D_model_independent_reach} shows that even FASER probes new parameter space for $c \tau_\phi / m_\phi \lesssim 0.3$.  
Not surprisingly, FASER2 does even better, probing new parameter space for $c \tau_\phi / m_\phi \lesssim 0.1$, with a sensitivity to branching ratios $\text{Br} ( D \to \pi \phi ) \cdot \text{Br}(\phi \to \text{Visible})$ that are many orders magnitude smaller than current bounds.  

We stress that some caution must be used when interpreting a plot like  \Fig{fig:B_model_independent_reach}.
First, different experiments are typically sensitive to different final states of $\NPfield$. In particular, LHCb and NA62 are only sensitive to a $\mu^+ \mu^-$ final state, and therefore vanish for $m_\NPfield < 2m_\mu$, while CHARM was only sensitive to final states containing $e^+e^-, \mu^+\mu^-$, and $\gamma \gamma$.
Second, one should also keep in mind that, although it is not visible in \Fig{fig:B_model_independent_reach}, some experiments have blind spots. 
For example, the LHCb constraint shown in~\Fig{fig:B_model_independent_reach}, which is a search for prompt decays, unavoidably loses sensitivity for values of $m_\NPfield$ close to known QCD resonances in the relevant mass range, \eg, around $m_\NPfield \approx m_{J/\psi} \simeq 3.1~$GeV.

Any specific model would typically appear as a straight line in the planes of \Fig{fig:B_model_independent_reach}, \Fig{fig:D_model_independent_reach}, where each point on the line fixes the coupling strength of $\NPfield$ to the SM. 
Decreasing the coupling strength makes $\NPfield$ more weakly-coupled and longer-lived, \ie, one finds smaller branching ratios at higher values of $\tau_\NPfield$. 
If we allow $\NPfield$ to have another coupling controlling its decay to a dark sector, than a model line could deviate from a straight line at sufficiently small coupling to the SM, since the branching ratios could vanish at some finite $\tau_\NPfield$, where the $\NPfield$ predominantly decays to the dark sector.

If  $\NPfield$  promptly decays to hadrons inside the detector, the $\NPfield$ production rates could be constrained by measured 3-body hadronic decay rates of heavy mesons, including in particular $B\to K\pi\pi$ decays, which were measured at BaBar~\cite{BaBar:2009jov} and Belle~\cite{Belle:2006ljg}.
However, for $B \to K\phi \to K \pi\pi$ to contribute to the measured rates, the $\NPfield$ decay length in the lab  must be smaller than the vertex-tracking resolution of the detector. 
The spatial resolution of these detectors ranges from approximately 
$10^{-5}$~m to $10^{-3}$~m~\cite{BaBar:2001yhh,Belle:2000cnh}, depending on the detector and the specific region. 
This resolution is smaller than the $\NPfield$ decay lengths we consider in this work. 
Therefore, decay channels involving $\phi$ are unlikely to be selected in the experimental analyses, and these measurements do not constrain our parameter space.

%%%%%%%%%%%%%%%%%%%%%%%%%%%%%%%%%%%%%%%%%%%%
%%%%%%%%%%%%%%%%%%%%%%%%%%%%%%%%%%%%%%%%%%%%%%
\section{Flavored Scalar Models}
\label{sec:models}
%%%%%%%%%%%%%%%%%%%%%%%%%%%%%%%%%%%%%%%%%%%%
%%%%%%%%%%%%%%%%%%%%%%%%%%%%%%%%%%%%%%%%%%%%%%

The $\NPfield$ couplings to the SM can be parametrized model independently by the dimension-five effective field theory (EFT) Lagrangian,
\begin{align}\label{dim5}
    \mathcal{L}_\NPfield = {\cal L}_{\NPfield\bar{f}f}+ {\cal L}_{\NPfield VV}+{\cal L}_{\NPfield H}\,,
 \end{align}
 where
\beq
    {\cal L}_{\NPfield\bar{f}f} =   \frac{C_{ij}^u}{\Lambda} \tilde H \NPfield \bar{Q}_i U_j 
+    \frac{ C_{ij}^d}{\Lambda} H \NPfield \bar{Q}_i D_j 
+    \frac{ C_{ij}^\ell}{\Lambda} H \NPfield \bar{L}_i E_j
+ \text{h.c.}\,, \label{eq:Lff}
\eeq
and
\beq
   {\cal L}_{\NPfield VV} =  \frac{c_{gg}}{\Lambda} \frac{\alpha_s}{4\pi}\NPfield G^{\mu\nu}  G_{\mu\nu} +
    \frac{c_{WW}}{\Lambda} \frac{\alpha_2}{4\pi}\NPfield W^{\mu\nu}  W_{\mu\nu} +
    \frac{c_{BB}}{\Lambda} \frac{\alpha_1}{4\pi}\NPfield B^{\mu\nu}  B_{\mu\nu} \label{eq:LVV}\,.
\eeq
Here $H$ is the SM Higgs doublet, $Q$ are the SU(2) doublet quarks, $U$ and $D$ are the up and down-type SU(2) singlet quarks, and $L$, $E$ are the SU(2) doublet and singlet leptons, respectively.  
$\Lambda$ denotes the EFT scale at which the $\NPfield$ couplings are generated, and $C^{f}_{ij}$ are dimensionless $3\times3$ matrices of Wilson coefficients, where $f=\{u,d,\ell\}$ labels the fermion species.  
We assume that  the couplings are real, such that $\phi$ can be taken to be CP-even. 
The Lagrangian $ {\cal L}_{\NPfield H}$ contains $\NPfield$-Higgs couplings, which could generically induce $\NPfield$-Higgs mixing and a $\NPfield$ mass.

The relative sizes of the Wilson-coefficients $C^u$, $C^d$, and $C^\ell$ depend on the UV physics at $\Lambda$.  
Thus, \eg, heavy vector-like fermions that only have tree-level couplings to SM up-type quarks would generate the couplings $C^u$ at tree-level, with all other couplings generated at one-loop or higher. 
In contrast, models featuring heavy fermions with tree-level couplings to all SM fermions lead to $C^u$, $C^d$, and $C^\ell$ that are roughly comparable.  
Below, we refer to the former as \emph{FNU models} and to the latter as \emph{FN models}.  
Similarly, the sizes of Higgs-$\NPfield$ couplings in $ {\cal L}_{\NPfield H}$ depend on the UV completion.  
We assume that this UV completion does not generate marginal $\NPfield$-Higgs couplings, such that, in the low-energy theory, $\NPfield$ is light, and its mixing with the Higgs is mainly induced by the couplings of \Eq{eq:Lff}.\footnote{Without additional model building, this clearly involves fine-tuning, since terms like $\NPfield^2 H^\dagger H$ cannot be protected by symmetries. This problem is common to many light-scalar scenarios, and we do not address it here. Neither do we address the Higgs naturalness problem.}

Below the electroweak breaking scale, the $\NPfield$ couplings to fermions are given in the fermion mass basis by
 \beq
    {\cal L}_{\NPfield\bar{f}f} =   \frac1{\sqrt{2}}\,\varepsilon \left[c_{ij}^u  \NPfield \, \bar{u}_i P_R u_j 
+    c_{ij}^d  \NPfield \, \bar{d}_i P_R d_j
+      c_{ij}^\ell  \NPfield \, \bar{\ell}_i P_R \ell_j
+ \text{h.c.}\right]\,, \label{eq:Lff_after_EWSB}
\eeq
where $\varepsilon \equiv v/\Lambda$. 
The fermion-mass basis couplings are
\begin{equation}
 c^f\equiv    U^f_L
  C^f 
  {U^f_R}^\dagger\,,
\end{equation}
where $U^f_{L/R}$ are the unitary matrices that diagonalize the SM Yukawa matrices, such that
\begin{equation}
U^f_L\, Y^f \,  {U^f_R}^\dagger = \hat{Y}^f\,,
\end{equation}
is diagonal, and the CKM matrix is given by $V \equiv V_L^u V_L^{d \dagger}$. 
The equivalent matrices in the lepton sector account for neutrino mixing. 
We return to these below. 

The $\NPfield$ couplings to vector bosons take the form
\beq
   {\cal L}_{\NPfield VV} =  \frac{c_{gg}}{\Lambda}\frac{\alpha_s}{4\pi} \NPfield G^{\mu\nu}  G_{\mu\nu} +
    \frac{c_{\gamma\gamma}}{\Lambda}\frac{\alpha}{4\pi} \NPfield F^{\mu\nu}  F_{\mu\nu} +... \label{eq:LVV_after_EWSB}\,,
\eeq
where $c_{\gamma\gamma}= c_{BB} +  c_{WW}$, and we neglect the couplings to $W$ and $Z$ bosons.

Given our assumptions above, $\phi$-Higgs mixing is predominantly generated by the operators of \Eqs{eq:Lff_after_EWSB}{eq:LVV_after_EWSB}, so that
 $\theta\lesssim v/\Lambda$.
In~\App{app:higgs_mixing} we discuss the contributions of ``tree-level'' $\NPfield$-Higgs mixing.
As we show, these contributions only modify the order-one numbers in the diagonal entries
of $c^f$, so that they
are automatically taken into account in our analysis.
The mixing also generates finite contributions to the off-diagonal couplings at one-loop. For these to be sub-dominant, the scale $\Lambda$ is constrained to be lower than $10^{14}-10^{17}$~GeV, depending on the model.
Conversely, in dark Higgs models with $\sin\theta=10^{-4}$, the contributions of higher-dimension operators can be neglected only for $\Lambda\gtrsim10^{17}$~GeV
in generic models. 
In models  with extra flavor structure, such as the flavored models, this is relaxed to $\Lambda\gtrsim10^{14}$~GeV.

The experimental signatures of $\NPfield$ crucially depend on the flavor structure of the Lagrangian of \Eq{dim5}. 
For $\NPfield$ masses in the range
$m_K-m_\pi<m_\phi<m_B-m_K$, which we focus on, two dominant production channels are $D$- and $B$-meson decay.
Note that at lower masses, the FASER sensitivity to kaon decays is typically comparable to $B$ decays~\cite{Feng:2017vli}.
Assuming the MFV ansatz, the coupling matrices $C^u$, $C^d$, and $C^\ell$ are proportional, at leading order in the spurion expansion, to the SM Yukawa matrices, such that the mass-basis couplings $c^f$ are diagonal, and flavor-violating processes are suppressed.
The implications for $\NPfield$ production are dramatic. 
Since flavor-violating $\NPfield$ couplings are only generated by weak loops, $\NPfield$ production from down-type mesons is dominant~\cite{Bezrukov:2009yw,Boiarska:2019jym,Feng:2017vli,Kling:2022uzy}.
This can be seen by calculating the ratio of couplings in the MFV scenario using a spurion analysis,
\begin{align}
\left|\frac{c^d_{sb}}{c^u_{uc}}\right|_{\text{\tiny MFV}}
    \sim  \left|\frac{[V^\dagger (\hat{Y}^u)^2 V \hat{Y}^d]_{sb}}{[V (\hat{Y}^d)^2 V^\dagger \hat{Y}^u]_{uc}} \right| \sim \left|\frac{V^*_{ts} y_t^2V_{tb} y_b }{V_{ub}y_b^2 V^*_{cb}y_c  } \right| 
    \sim \left|\frac{1  }{y_b y_c V_{ub} } \right| 
    \sim 10^6\,.
    \label{eq:MFV_couplings}
\end{align}

While there are models that realize the MFV ansatz, most notably dark Higgs 
models~\cite{Patt:2006fw,Beacham:2019nyx}, 
the small flavor mixing in the quark sector suggests a simple and elegant alternative for controlling $\NPfield$-mediated flavor-violating processes, based on the \emph{alignment}~\cite{Nir:1993mx} of new flavor parameters with the Yukawas~\cite{Nir:1993mx,Leurer:1993gy}.
 The main observation in alignment models is that any new physics that explains the SM fermion flavor structure also controls the textures of the new physics flavor structure, in our case, the $\NPfield$-fermion couplings.
As a concrete example, we take the simplest prototype flavor model, namely a single $U(1)$ Froggatt-Nielsen (FN) flavor symmetry~\cite{Froggatt:1978nt}, with an appropriate choice of fermion charges, \ie, one that generates the observed SM flavor hierarchies.
Various such choices have been considered in the literature; see, \eg, Refs.~\cite{Leurer:1992wg,Leurer:1993gy}.
The only freedom, then, is in the $U(1)$ charge of $\NPfield$. 
Here we make the simplest choice and take $\NPfield$ to be neutral under the flavor symmetry.
$C^f$ and the SM Yukawas $Y^f$ then have
the same texture in the interaction basis.
In the mass basis, $\hat{C}^f$ inherits the same texture.
In particular, in the quark sector we have the following flavor textures,
\begin{gather} 
   c^u
   \sim
    \begin{pmatrix}
    \lambda^6 & \lambda^4 & \lambda^3\\
    \lambda^5 & \lambda^3 & \lambda^2\\
    \lambda^3 & \lambda & 1
    \end{pmatrix}
    \,, \;\;
      c^d
      \sim
\begin{pmatrix}
    \lambda^6 & \lambda^5 & \lambda^5\\
    \lambda^5 & \lambda^4 & \lambda^4\\
    \lambda^3 & \lambda^2 & \lambda^2
    \end{pmatrix}
    \label{eq:cu_cd_texture}\,,
\end{gather}
where $\lambda$ is the FN symmetry breaking parameter, which is commonly taken to be the Cabibbo mixing, or $\lambda=0.2$.
 As for the lepton sector, a single $U(1)$ symmetry cannot account for the hierarchical charged lepton masses and large neutrino mixings~\cite{Grossman:1998jj}. 
 These can be explained, \eg, by a discrete symmetry or two $U(1)$s, with a range of possibilities for the off-diagonal entries of $Y^\ell$~\cite{Grossman:1998jj,Feng:1999wt}, and therefore also of $C^\ell$. 
Since the latter have little effect on the phenomenology, we take $c^\ell$ to be proportional to the SM lepton Yukawa matrix, that is,
\beq\label{eq:cl}
c^\ell=\text{diag}(m_e,m_\mu,m_\tau)/v.
\eeq

The ``$\sim$'' sign in \Eq{eq:cu_cd_texture} indicates that these equations merely display the scaling of the different entries with $\lambda$.
Each entry involves an unknown $\mathcal{O}(1)$ prefactor. 
Thus, \Eq{eq:cu_cd_texture} defines a
broad class of models with a range of possible low-energy phenomenologies which depend on the various unknown $\mathcal{O}(1)$ prefactors. 
In particular, the $\NPfield$ lifetime and the different production and decay rates may vary by one or two orders of magnitude.
Note that the flavor-diagonal $\NPfield$ couplings in $c^u$ and $c^d$ have the same $\lambda$ scaling as the SM Yukawa couplings, 
\begin{align}
    y_u \sim y_d \sim y_c^2 \sim \lambda^6 \quad\quad\text{and}\quad\quad y_s\sim y_b^2 \sim \lambda^4\,,
\end{align}
just as in MFV models.
Importantly however, unlike in MFV models, the off-diagonal entries of $c^f$ can be nonzero.

To summarize, we consider two types of models:
\begin{enumerate}
\item FN models: The matrices $C^u$, $C^d$, and $C^\ell$ are all generated at tree-level at $\Lambda$. The $\NPfield$ couplings  in the fermion mass basis, $c^u$, $c^d$, $c^\ell$ are  given by \Eqs{eq:cu_cd_texture}{eq:cl}, and 
\begin{align}
   \left|\frac{c^d_{bs}}{c^u_{uc}}\right|_{\text{\tiny FN}}
    \sim  \frac{1}{\lambda^2} 
    \sim 10^2\,.
    \label{eq::FN_couplings}
\end{align}
\item FNU models: Only $C^u$ is generated at tree-level at $\Lambda$. 
The $\NPfield$ coupling to up quarks  in the fermion mass basis,
$c^u$, is  given by \Eq{eq:cu_cd_texture}, 
with $c^d=c^\ell=0$ at tree-level.
\end{enumerate}
In the FN model, in experiments in which $N_D \gtrsim 10^4 N_B$, such as the CHARM experiment, $\NPfield$ production via rare $D$-meson decays would be comparable to or larger than production from $B$-meson decays.
In the FNU model, a smaller hierarchy between $N_D$ and $N_B$ is sufficient for the two production channels to be comparable.

Flavor-changing couplings in the down-sector in the FNU models are unavoidably generated by the weak interactions. 
Integrating out the $W$ boson at one loop, we find the matching contribution
    \begin{align}
        \Delta c^d&= \frac{\varepsilon}{\sqrt{2}} \frac{V^\dagger 
        [3c^u {(\hat{Y}^u)}^\dagger+5\hat{Y}^u {(c^{u})}^\dagger ]V\hat{Y}^d}{16\pi^2} 
        \log \left( \frac{\Lambda^2}{      \mu_{\text{\tiny IR}}^2}\right) \,,
        \label{eq::FNu_bottom_coupling}
    \end{align}
with $\mu_{\text{\tiny IR}} \sim \mathcal{O}(1~\text{GeV})$.
Note that, in contrast to dark Higgs models, the 1-loop contribution to the flavor-violating couplings is UV sensitive. 
In this case, we find
\begin{align}
   \left|\frac{c^d_{sb}}{c^u_{uc}}\right|_{\text{\tiny FNU}}
    \sim  \frac{3}{16\pi^2} \frac{V^*_{cs}  V_{tb} y_t y^2_b}{y^2_b}\log \left( \frac{\Lambda^2}{      \mu_{\text{\tiny IR}}^2}\right)\sim 10^{-2}\log \left( \frac{\Lambda^2}{      \mu_{\text{\tiny IR}}^2}\right)\,,
    \label{eq:FNu_coupling_ratio}
    \end{align}
which is smaller than the ratio in \Eq{eq::FN_couplings}, even for UV scales as high as the Planck scale. Varying the prefactors in \Eq{eq:cu_cd_texture} would lead to $\mathcal{O}(1)$ variation of the coefficient of the logarithmic factor in \Eq{eq:FNu_coupling_ratio}.

%%%%%%%%%%%%%%%%%%%%%%%%%%%%%%%%%%%%%%%%%
\section{Flavored Scalar Signatures at FASER}
\label{sec:pheno}
%%%%%%%%%%%%%%%%%%%%%%%%%%%%%%%%%%%%%%%%%

We now turn to the production and decay of $\NPfield$ particles in the flavored models.
These predictions depend, of course, on the unknown order-one prefactors in the matrices $c^f$.  
We will show the resulting variations in the $\NPfield$ lifetime and muon branching fraction, but for the most part, focus on a particular benchmark point. 
Our analysis closely follows Ref.~\cite{Kling:2022uzy}, which studied the FASER reach for MFV scalars.  
The relative importance of $\NPfield$ production from $B$-meson and $D$-meson decay can be estimated using parton-level amplitudes up to hadronic uncertainties, which are typically small for heavy mesons. 
The $D$-meson branching fraction is, at leading order in $\mathcal{O}(|c^u_{uc}|/|c^u_{cu}|)$,
    \begin{align}
        &\text{Br}(D \to \pi \; \phi)   \approx \frac{\varepsilon^2m_{c}|c^u_{uc}|^2}{32\pi \Gamma_{D^0}}F\left[\frac{m_u}{m_c},\frac{m_\phi}{m_c}\right]\approx 
10^{-11}\left( \frac{10^{10}\,\text{GeV}}{\Lambda}\right)^2\left( \frac{|c^u_{uc}|}{\lambda^4}\right)^2 \left(\frac{F}{1}\right)\,,
    \end{align}
where $\Gamma_{D^0}$ denotes the SM width, and $F\left[x,y\right]\equiv \sqrt{1-(x+y)^2}\sqrt{1-(x-y)^2}(1+x^2-y^2)$. 
Similarly, for the $B$ meson, 
    \begin{align}
        \text{Br}(B \to K \; \phi) &\approx   \frac{\varepsilon^2m_{b}|c^d_{bs}|^2}{32\pi \Gamma_{B^0}}F\left[\frac{m_s}{m_b},\frac{m_\phi}{m_b}\right]\approx 10^{-7}\left( \frac{10^{10}\,\text{GeV}}{\Lambda}\right)^2\left( \frac{|c^d_{bs}|}{\lambda^2}\right)^2 \left(\frac{F}{1}\right)\,,
    \end{align}
where $\Gamma_{B^0}$ denotes the SM width.
As expected, for FN models, 
these partial widths differ by four orders of magnitude 
due to the hierarchy of the couplings seen in~\Eq{eq::FN_couplings}.

The optimal search strategy for $\NPfield$ strongly depends on its typical propagation length in the lab, as well as on its visible final states and their corresponding branching fractions.
In the mass range of interest, the scalar decays primarily to hadrons, with subleading decays to muons and/or photons in parts of the model parameter space.
Despite significant efforts to estimate the hadronic decay rates, substantial theoretical uncertainties persist in their values. 
For a recent discussion, see Ref.~\cite{Blackstone:2024ouf} and references therein.
For $m_\phi\lesssim 2\,$GeV, we calculate the hadronic rates using hadronic form factors, and 
for $m_\phi\gtrsim 2\,$GeV we use the perturbative spectator model.
These are adequate for the estimates we present here, because of the inherent large variations in the model predictions due to the unknown numerical prefactors in the Wilson coefficients.
For more details on the calculations of the rates, see \App{app:decays}.

%%%%%%%%%%%
\begin{figure}[t]
    \centering    \includegraphics[width=0.49\textwidth]{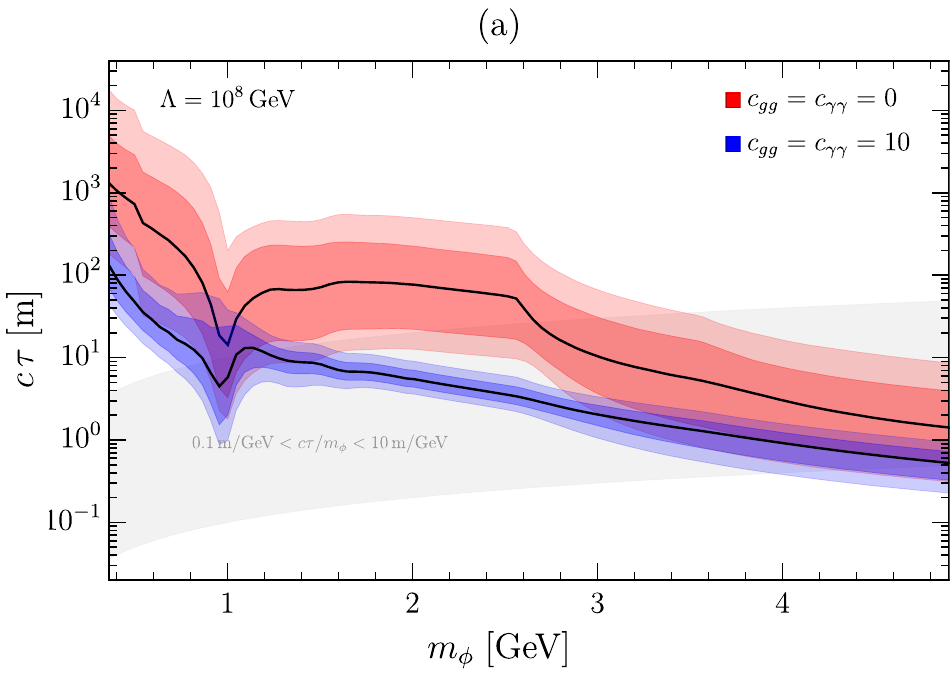} 
    \hfill
    \includegraphics[width=0.49\textwidth]{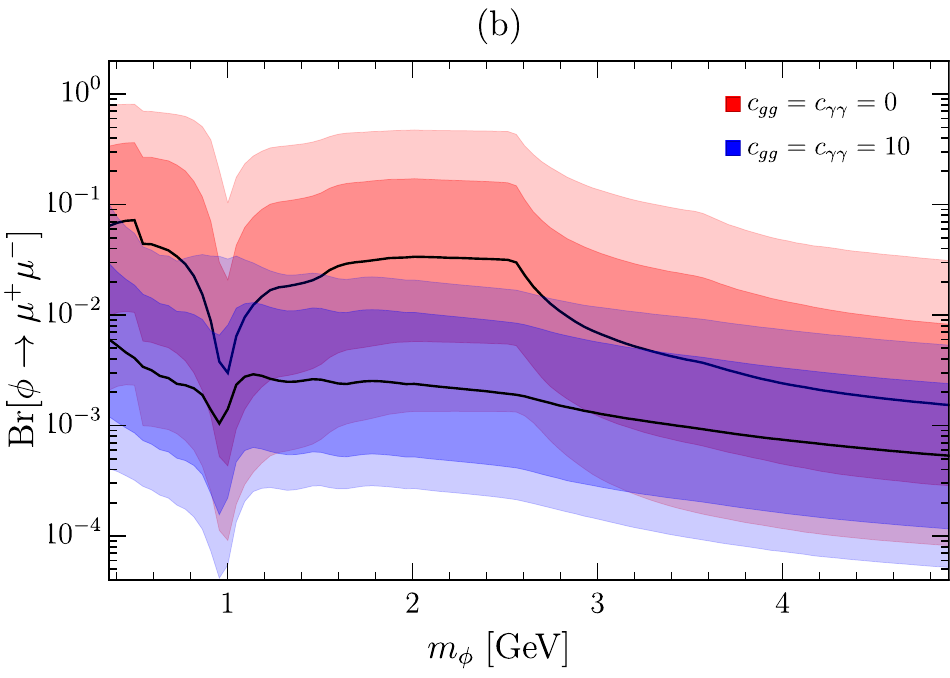}
    \caption{Variations in the $\NPfield$ lifetime and dimuon branching fraction in the FN models.  Left: lifetime as a function of $m_\NPfield$ for the gluophobic (-philic) case in red (blue) for $\Lambda=10^8\,$GeV. The bands represent the $1\sigma$ and $2\sigma$ widths due to $[1/3-3]$ variations of all the diagonal scalar coupling around their SM Yukawa values with arbitrary sign, with the medians plotted as black solid lines. For reference, the peak sensitivity region of FASER is plotted in gray. Right: Same for branching fraction to dimuons. 
    }  
    \label{fig:ctau_BRs}
\end{figure}
%%%%%%%%%%%

We show the model predictions for the $\NPfield$ lifetime and dimuon branching ratio as functions of the $\NPfield$ mass for a fixed EFT scale $\Lambda$ in \Fig{fig:ctau_BRs}, where we consider both 
gluophobic models ($c_{gg}=0$) and 
gluophilic models ($c_{gg}=10$) (see \Eq{eq:gluon_eff}). 
The off-diagonal couplings have little effect on the lifetime and are neglected here.
To obtain these results, we scan over different values of the prefactors in the $\NPfield$-fermion couplings $c^f$.
Concretely, we randomly choose the prefactor for each diagonal entry $c^f_{ii}$
in the range $(1/3-3)\, y_i^{\text{\tiny SM}}$ with arbitrary signs, using a uniform probability distribution on a log scale.
To get a sense for the effects of these order-one prefactors, we plot 
the ``$1\sigma$'' and ``$2\sigma$'' ranges of the resulting distributions for each mass, with the medians shown as solid black curves.
As expected, the total lifetime scales as $c\tau \propto\Lambda^2$ and is smaller in the gluophilic models compared to the gluophobic models.
In both cases, the lifetime is dominated by the hadronic channels.
In the gluophobic models, the order-one variations in the $\NPfield$ couplings to fermions lead to variations of roughly one order of magnitude in the $\NPfield$ lifetime for a given mass.
These can significantly affect the experimental reach for the models, which is highly sensitive to the total lifetime. 
The variations in the lifetime become milder in the gluophilic models because of the larger coupling to gluons.
The gray band shows the region of peak sensitivity of FASER and FASER2, roughly defined by $0.1 \,\text{m}/\text{GeV}<c\tau /m<10 \,\text{m}/\text{GeV}$; see, \eg, \Fig{fig:B_model_independent_reach}. 

We also show the analogous variations in 
the residual rate to muons in the right panel of \Fig{fig:ctau_BRs}.
This is the dominant non-hadronic rate below the $\tau^+ \tau^-$ threshold. 
The bands are calculated in the same manner as in the left panel.
Importantly, the larger coupling to gluons leads to an average suppression of the leptonic channel from $\mathcal{O}(1\%-10\%)$ in the gluophobic case to $\mathcal{O}(0.1\%)$ in the gluophilic case.
The dominant hadronic channel varies with the $\NPfield$ mass.
For $m_\phi\lesssim 1\,$GeV, the width is dominated by the $\pi \pi$ channel, for $1\,\text{GeV}\lesssim m_\phi\lesssim1.5\,$GeV it is dominated by the $KK$ channel, and for $1.5\,\text{GeV}\lesssim m_\phi \lesssim 2\,$GeV it is dominated by the remaining hadronic channels including $4\pi$, $\eta\eta$, etc.
In the gluophobic models, the dominant  channels are $\bar{s}s$ and $gg$ for $m_\phi \gtrsim 2\,$GeV until the $\bar{c}c$ channel becomes kinematically accessible and dominant.
In the gluophilic case, the $gg$ channel dominates throughout.
The values of these separate branching fractions do not affect our
results for FASER below, since, as explained in Section~\ref{sec:modindep}, we assume $\epsilon=1$ efficiency for all SM decays at FASER. 
For some of the other searches we consider, $Br(\NPfield\to\mu\mu)$, which is the only appreciable non-hadronic channel, is also relevant.

\begin{figure}[tbp]
    \centering    \includegraphics[width=0.49\textwidth]{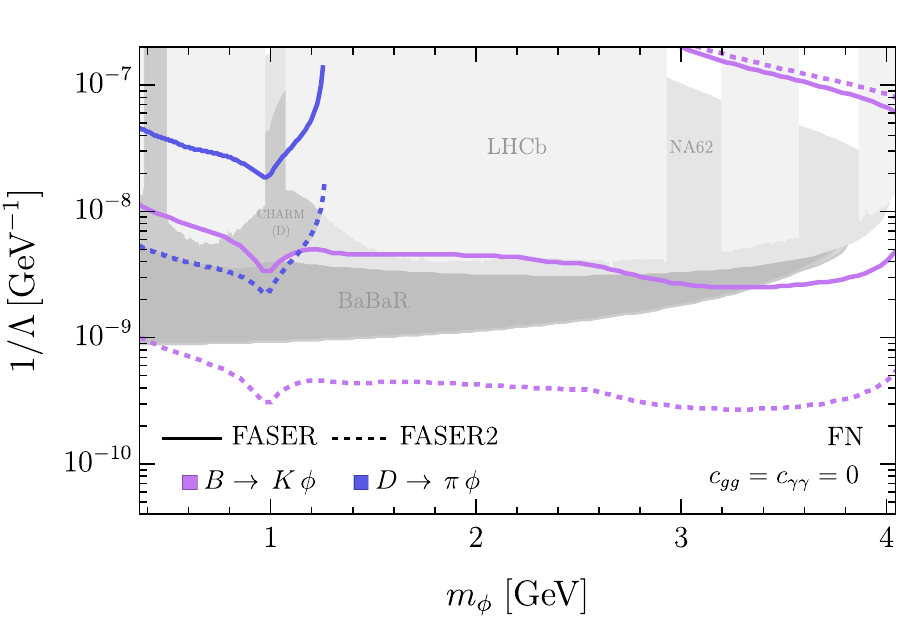} 
    \hfill
    \includegraphics[width=0.49\textwidth]{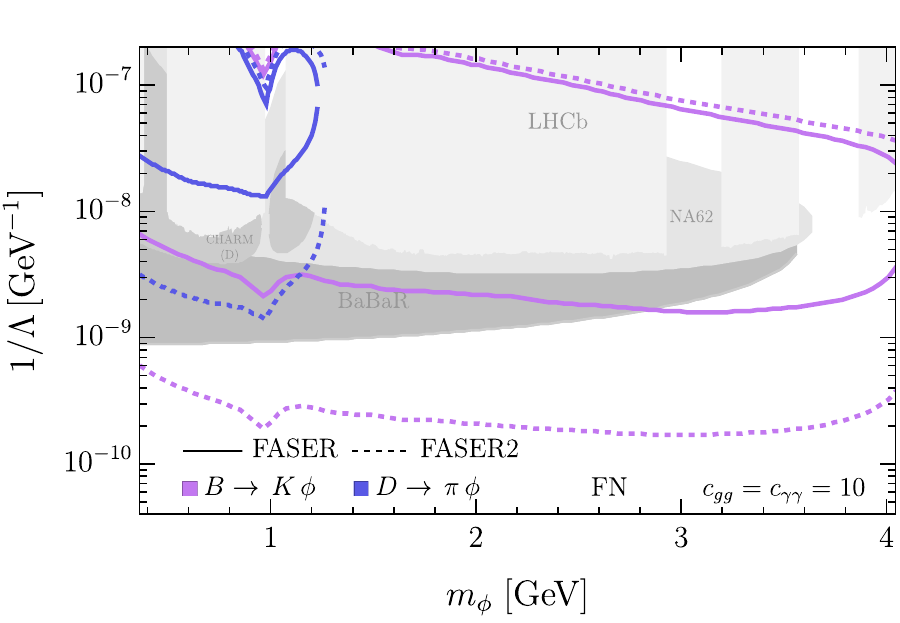}
    \caption{Left: projected sensitivity for FASER and FASER2 in solid and dashed lines, respectively, from $B$ ($D$) meson decays in purple (blue) for the gluophobic ($c_{gg}=0$) FN model. In gray we plot existing constrains from LHCb~\cite{LHCb:2015nkv,LHCb:2016awg}, CHARM~\cite{CHARM:1985anb} and NA62~\cite{NA62:2023qyn}, which are sensitive only to the $\mu^+ \mu^-$ final state.    
    We also plot the constraints due to the $B\to K\,+\text{inv.}$~interpretation~\cite{MartinCamalich:2020dfe} of the BaBar $B \to K\,\bar{\nu}\nu$ search~\cite{BaBar:2013npw} using \Eq{eq:BaBar_search}. Right: same as in the left panel, but for for the gluophilic ($c_{gg}=10$) FN model. 
    }  
    \label{fig:FN_models}
\end{figure}

The FASER sensitivity to $\NPfield$ particles in FN models is shown schematically in \Fig{fig:FN_models}.
The purple curve is obtained by considering $\NPfield$ particles from $B$ decays only, while the blue curve corresponds to $\NPfield$'s from $D$ decays only. 
We consider a benchmark FN model with $c^u$ and $c^d$ given by~\eqref{eq:cu_cd_texture} with all prefactors taken to be 1, and with $c^\ell$ given by~\eqref{eq:cl}. 
We have verified that different benchmarks, obtained by varying the prefactors, lead to reasonable variations, with no accidental cancellations.

As expected, we find that, as opposed to MFV models, both $B$- and $D$-meson decays are efficient probes of the FN model, although $B$-meson decay still provides an overall stronger probe; it probes higher UV scales due to the 
couplings hierarchy of \Eq{eq::FN_couplings} and a larger portion of parameter space where $D$ meson decays are kinematically forbidden. 
The gray regions show existing experimental constraints from LHCb~\cite{LHCb:2015nkv,LHCb:2016awg}, CHARM~\cite{CHARM:1985anb}, and NA62~\cite{NA62:2023qyn}.\footnote{$D$-$\bar{D}$ mixing provides a weaker bound, which we estimate to be $\Lambda \gtrsim 10^6~$GeV, adapting the approach of Ref.~\cite{Carmona:2021seb}.} 
The NA62 bounds were extrapolated to smaller and larger masses by assuming the bounds reported in Ref.~\cite{NA62:2023qyn} depend only on $c\tau_\NPfield/m_\NPfield$. 
For larger masses this is a conservative estimate, since the sensitivity may slightly improve thanks to the improved geometrical acceptance, as shown in Section~\ref{sec:modindep}. 
We conservatively assumed that CHARM is only sensitive to final states containing a lepton- or photon-pair. 
In principle, hadronic final states may be observable at CHARM with some efficiency, for instance two neutral pions decaying promptly to four photons.
The estimation of the CHARM sensitivity to such fully hadronic final states is nontrivial, and we leave it for future work.
We also plot the constraints due to the $B\to K\,+\text{inv.}$ interpretation~\cite{MartinCamalich:2020dfe} of the BaBar $B \to K\,\bar{\nu}\nu$ search~\cite{BaBar:2013npw}, using \Eq{eq:BaBar_search}. 
We find that in the gluophobic case, FASER2 covers unexplored parameter space in which $\NPfield$ is too short lived to be ruled out by BaBar and the decay to muons is sufficiently suppressed to make it out of reach for CHARM and NA62.
In the gluophilic case, the sensitivities of CHARM, NA62, and BaBar are degraded at lower scales due to the shorter $\NPfield$ lifetime.  
CHARM and NA62 also lose sensitivity at high scales due to the decreased branching ratio to muons.  
The FASER reach is sensitive only to the increased lifetime, which causes the sensitivity region to shift downwards to higher UV scales. 

\begin{figure}[tbh]
    \centering    \includegraphics[width=0.49\textwidth]{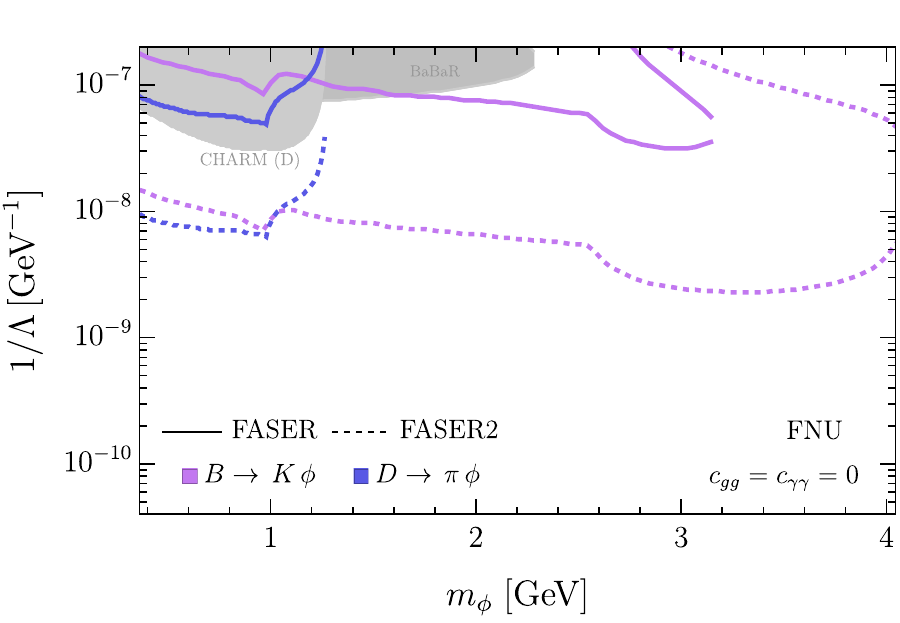} \includegraphics[width=0.49\textwidth]{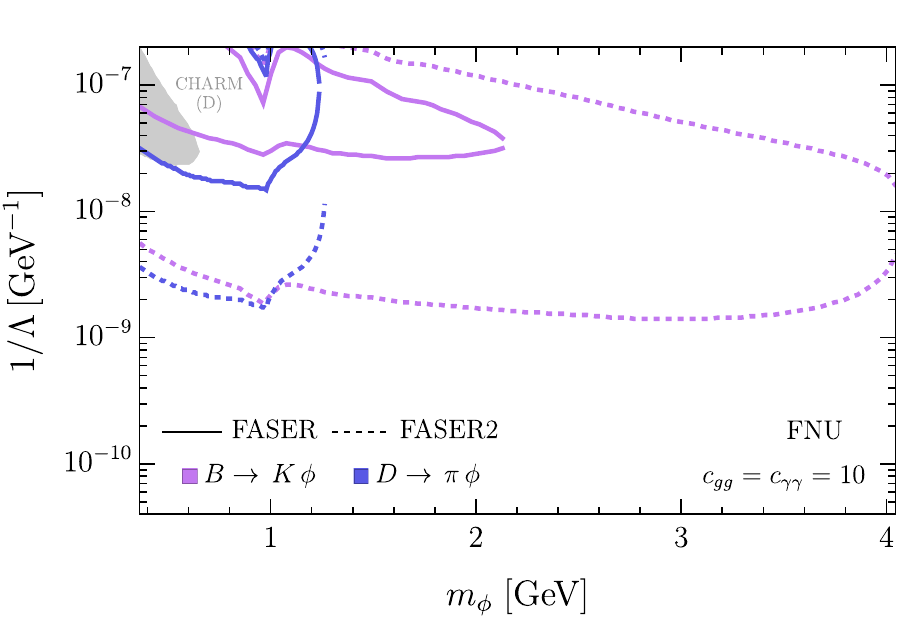}
    \caption{Same as in \Fig{fig:FN_models}, but for the FNU model.  }  
    \label{fig:FNu_models}
\end{figure}

Similar plots are shown in \Fig{fig:FNu_models} for the FNU model.
The benchmark considered here has $c^\ell=c^d=0$ and $c^u$  given by~\eqref{eq:cu_cd_texture}, with all prefactors taken to be 1, as above.
One main difference is the reduced rate for $B$ decays due to the suppression of the coupling, for which we use the approximation in \Eq{eq:FNu_coupling_ratio}.
This leads to reduced sensitivity for all the probes of $B$-meson decay like BaBar,
while the FASER sensitivity from $D$-meson decays is relatively unaffected and shifts only due to the change in lifetime.
In addition, since $c^\ell=0$, the LHCb and NA62 bounds are no longer relevant, and the CHARM bounds from $D$ decays are still present due to the decay rate to photons.
In the FNU model, $D$ decay proves to be a comparable if not slightly stronger probe than $B$ decay in the kinematically allowed region, in stark contrast to MFV models.
In addition, in both the gluophobic and the gluophilic cases, FASER is left as a unique probe in large parts of parameter space due to its sensitivity to hadronic final states.

It should be stressed that the sensitivities shown in the plots are fairly crude estimates, based on $\NPfield$ production from \emph{either} $B$ \emph{or} $D$ decays. In practice, the two sources will contribute, leading to a somewhat enhanced sensitivity. Importantly, in some of the regions accessible to FASER, and in particular FASER2, FNU models predict comparable numbers of $\NPfield$ from these two sources. In FN models, $B$ decays dominate,
but $D$ decays also contribute. 
Since $\NPfield$ decays are fully reconstructible at FASER, 
it may be possible to distinguish the two.
With enough statistics, 
one can form doubly differential distributions as functions of distance from the beam collision axis and energy, and disentangle the $\NPfield$ production mechanisms.  A proof of principle of this approach has, in fact, already been demonstrated, using $\sim 300$ neutrino events at FASER~\cite{FASER:2024ref}, where the energy distribution has been used to determine the relative proportion of events from $\pi$ and $K$ decay.  
%%%%%%%%%
\begin{figure}[t]
    \centering    \includegraphics[width=0.8\textwidth]{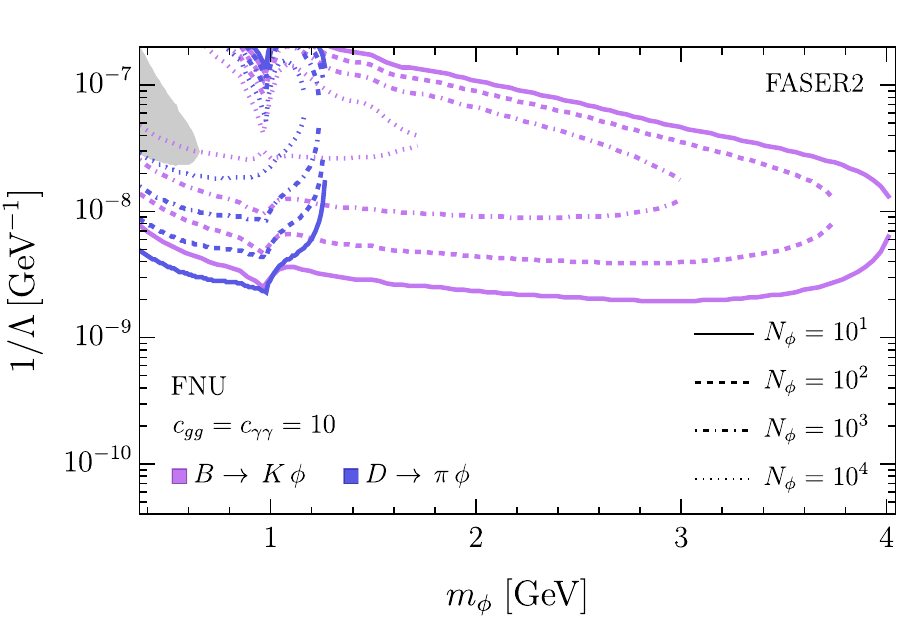} 
    \caption{Contours of number of events in FASER2 in the gluonphilic FNU model. Events from $B$ ($D$) decays are shown in purple (blue) with solid, dashed, dot-dashed, and dotted curves for $10, 10^2, 10^3$ and $10^4$ signal events, respectively.
    The gray region is excluded by $D$ decays in CHARM.}   \label{fig:event_contour}
\end{figure}
%%%%%%%%%
This requires sufficient statistics, but as shown in \Fig{fig:event_contour}, there are regions of parameter space where hundreds or thousands of events could be detected at FASER2.  We leave it to future work to determine the extent to which events originating in $B$ and $D$ decays can be differentiated, but such results would be extremely interesting.  For one, they would establish whether the new physics is MFV or non-MFV.  The latter would strengthen the case for a theory of flavor, such as
the simple flavor symmetry considered here, to keep flavor-violating processes under control, and so would provide new experimental handles on \emph{SM} flavor.

%%%%%%%%%%%%%%%%%%%%%%%%%%%%%%%%%%%%%%%%%%%%%%%%%%%%%%%%%%%%%%%%
\section{Conclusions and Outlook}
\label{sec:conclusions}
%%%%%%%%%%%%%%%%%%%%%%%%%%%%%%%%%%%%%%%%%%%%%%%%%%%%%%%%%%%%%%%%

We have studied the discovery reach of the FASER experiment and its planned FASER2 upgrade for flavored scalar LLPs.  These particles have generic couplings to the SM, subject to selection rules from an assumed flavor symmetry that generates the SM fermion masses and mixings.  They can be produced in either up-type or down-type heavy meson decays, and predominantly decay to hadrons.

Our analysis highlights several features of the FASER detector. First, because of the low backgrounds at FASER, it is sensitive to hadronic decays. Second, the decays of known hadronic resonances, which greatly degrade the sensitivity of near-IP detectors for  masses close to the resonance mass, are irrelevant at FASER, because of the large distance from the IP and the thick shielding provided by $\sim 100$ m of rock. Third, because of the large boosts of particles from LHC collisions, FASER can probe relatively small lifetimes compared with lower-energy experiments. 

 With these features as motivation, we have carried out a model-independent analysis of the experimental reach of FASER and FASER2 in terms of the LLP lifetime over mass ratio. This model-independent analysis yielded some very interesting conclusions. For example, for scalars produced in $B$ decays, FASER cannot probe new parameter space, but FASER2 is roughly a factor of $10^4$ more sensitive, and probes new parameter space for $c \tau_\phi / m_\phi \gtrsim 0.1$.  For scalars produced in $D$ decays, we have found that both FASER and FASER2 probe new parameter space for $c \tau_\phi / m_\phi \lesssim 0.1$ with a sensitivity that is many orders magnitude greater than current bounds.  In MFV models, the production of scalars from $D$ decays is typically suppressed, but this is highly model-dependent, and we have seen in this study that flavor models, where a symmetry explains the observed masses and mixings of the SM fermions, can naturally predict sizable production of flavored scalars from $D$ mesons. 

The simple model we study demonstrates the interplay between flavor and LLP models. From a phenomenological point of view, LLPs are characterized by small couplings to the SM, or they would not be long-lived. Longer lifetimes generically imply smaller couplings, which in turn allow for larger departures from MFV.
From a theory point of view, there is strong motivation
for some theory of flavor, which would give rise to the SM flavor structure.
Such a theory would generically give rise to a non-trivial flavor structure in LLP couplings, with striking implications for 
their searches.
Conversely, if LLPs exist and are within experimental reach, they will provide a host of new flavor measurements.

We have focused on a simple minimal setup, with a single $U(1)$ Froggatt-Nielsen flavor symmetry, and a $CP$-even scalar 
with zero $U(1)$ charge. There are various extensions of this basic model that would be interesting to explore. One involves more general flavor models, including different LLP flavor charges, extended flavor symmetries with non-trivial lepton couplings,  or altogether different theories of flavor. 
Another direction involves different  types of particles,
such as ALPs or spin-1 particles, with general non-renormalizable couplings to the standard model controlled by a flavor symmetry.

%%%%%%%%%%%%%%%%%%%%%%%%%%%%%%%%%%%%%%%%%%%%%%%%%%%%%%%%%%%%%%%%
\acknowledgments
We thank Michael Waterbury for collaboration at early stages of this project,
and Iftah Galon, Enrique Kajomovitz and Yotam Soreq for useful discussions. 
We are grateful to Babette D\"obrich and Jan Jerhot for their assistance with the recasting of the CHARM bounds, and to Iftah Galon for assistance with FONLL.
We also thank Yotam Soreq for comments on the draft.
Research supported by NSF-BSF (Grant No.~2020-785).  RB is supported by the U.S. Department of Energy grant number DE-SC0010107.
Part of the work by RB was performed at the Aspen Center for Physics, which is supported by National Science Foundation grant PHY-1607611.
NB and YS are also supported by the Israel Science Foundation (Grant No.~1002/23). 
The work of JLF was supported in part by U.S.~National Science Foundation Grants PHY-2111427 and PHY-2210283, Simons Investigator Award \#376204, Heising-Simons Foundation Grants 2019-1179 and 2020-1840, and Simons Foundation Grant 623683.

%%%%%%%%%%%%%%%%%%%%%%%%%%%%%%%%%%%%%%%%%%%%%%%%%%%%%%%%%%%%%%%%
%%%%%%%%%%%%%%%%%%%%%%%%%%%%%%%%%%%%%%%%%%%%%%%%%%%%%%%%%%%%%%%%
\appendix
%%%%%%%%%%%%%%%%%%%%%%%%%%%%%%%%%%%%%%%%%%%%%%%%%%%%%%%%%%%%%%%%

%%%%%%%%%%%%%%%%%%%%%%%%%%%%%%%%%%%%%%%%%%%%%%%%%%%%%
\section{$\NPfield$ Decays}
%%%%%%%%%%%%%%%%%%%%%%%%%%%%%%%%%%%%%%%%%%%%%%%%%%%%%%%%
\label{app:decays}

In this appendix, we collect the equations employed for the different decay channels, following Refs.~\cite{Winkler:2018qyg, Boiarska:2019jym, Kling:2022uzy, Boiarska:2019vid}.

\subsection{Photons}

The $\NPfield$ coupling to photons gets finite 1-loop contribution from all the charged states~\cite{Djouadi:2005gi,Kling:2022uzy},
\begin{align}
c_{\gamma\gamma}^{\text{eff}} &= c_{\gamma\gamma}+c_{\gamma\gamma}^{\text{1-loop}}\,,
    \\
 c_{\gamma\gamma}^{\text{1-loop}}&\equiv\sum_{f
    \in \{ u,d,\ell\}} \sum_{i\in f} N^f_c Q_i^2 \left( \frac{c^f_{ii}}{y_i}\right)A_{1/2}(\tau_i) \,,
\end{align}
where $N^f_c$ is the number of colors, 
$Q_i$ is the electric charge, and $\mathcal{A}_{1/2}$ is the loop function of the variable $\tau_i\equiv m_\NPfield^2/4m_i^2$. For its definition, see Sec.~2.3.1 of Ref.~\cite{Djouadi:2005gi}.
In the mass range of interest, $|c_{\gamma\gamma}^{\text{1-loop}}| \sim 3$. 
The decay to photons is then given by
\begin{align}
\Gamma(\NPfield\rightarrow\gamma\gamma) = \frac{m_\NPfield^3 \alpha^2 |c_{\gamma\gamma}^{\text{eff}}|^2}{64\pi^3 \Lambda^2} \,.
\end{align}
We find that for $c_{\gamma\gamma}\sim \mathcal{O}(1)$, the photon channel is subleading in the relevant mass range, since
\begin{align}
  &\Gamma(\phi \to \gamma \gamma)/\Gamma(\phi \to \mu^+ \mu^-) \approx 10^{-3}\,,
  \\
    &\Gamma(\phi \to \gamma \gamma)/\Gamma(\phi \to \text{hadrons}) \approx 10^{-5} \,.
\end{align}

\subsection{Leptons}

In the mass range of interest, the leptonic decay of the scalar is primarily to muons, 
\begin{align}
    \Gamma(\phi\to \ell\,\ell) 
            &=  \frac{\varepsilon^2m_\phi |c^\ell_{\mu\mu}|^2 \beta_\mu^3}{16\pi}\,, 
\end{align}
with $\beta_i \equiv \sqrt{1-4m_i^2/m_\phi^2}$, whereas the decays to $e\mu$ and $ee$ final states are suppressed by $(y_e/y_\mu)$ and $(y_e/y_\mu)^2$, respectively.

\subsection{Hadrons}

Below 2~GeV, the scalar decays to $\pi\pi$ and $KK$ with the following rates~\cite{Donoghue:1990xh,Winkler:2018qyg,Kling:2022uzy}:
\begin{align}   
\Gamma(\NPfield\to \pi \pi)&=\frac{3\beta_{\pi} }{32  \Lambda^2 \pi m_\phi} \left|
  \frac{6c^{\text{eff}}_{gg}}{27}(\Theta_{\pi}-\Gamma_{\pi}-\Delta_{\pi})+
\frac{c^u_{11}+c^d_{11}}{y_u+y_d} \Gamma_{\pi} + (c^d_{22}/y_s)\Delta_{\pi}
\right|^{2},
\label{eq:pipi_rate}
\\
\Gamma(\NPfield \to K K)&=\frac{ \beta_{K}}{8 \Lambda^2 \pi m_\phi} \left|
  \frac{6c^{\text{eff}}_{gg}}{27}(\Theta_K-\Gamma_K-\Delta_K)+
\frac{c^u_{11}+ c^d_{11}}{y_u+y_d} \Gamma_{K} + (c^d_{22}/y_s)\Delta_{K}
\right|^{2}\,.
\label{eq:KK_rate}
\end{align}
See, \eg, Ref.~\cite{Winkler:2018qyg} for the definition of the form factors $\Theta_{i},\Gamma_{i},\Delta_{i}$ for $i=\pi,K$, and the values used in this work. The effective coupling to gluons is 
\begin{align}
        c^{\text{eff}}_{gg} &= c_{gg} + c_{gg}^{\text{1-loop}}\,,
        \label{eq:gluon_eff}
        \\
        c_{gg}^{\text{1-loop}}&\equiv \frac12 \, 
    \sum_{f\in \{u,d\}}\sum_{i\in f} \left(\frac{c^f_{ii}}{y_i}\right)\mathcal{A}_{1/2}\left(\tau_f^\NPfield\right)\,,
\end{align}
where $c_{gg}^{\text{1-loop}}$ is generated at 1-loop due to the couplings to quarks.
In the mass range of interest, $|c_{gg}^{\text{1-loop}}| \sim 1$. 
We use the hadronic form factors calculated using dispersion relations~\cite{Winkler:2018qyg}, since the leading order $\chi$PT results already receive large corrections above $0.5\,$GeV.
For $1.3\,\text{GeV}<m_\phi<2~$GeV, additional hadronic final states, which are not covered by \Eqs{eq:pipi_rate}{eq:KK_rate}, become accessible.
We parameterize these additional contributions by~\cite{Winkler:2018qyg}
\begin{align}
    \Gamma(\NPfield\rightarrow 4\pi,\eta\eta,\rho\rho,...) \approx 9C_{4\pi}\varepsilon^2|c^{\text{eff}}_{gg}|^2 m_\NPfield^3 \beta_{2\pi} \,,
    \label{eq:hadrons_rest}
\end{align}
where $C_{4\pi}$ is a constant chosen such that the total hadronic rate is continuous across the transition at 2~GeV.
Above 2~GeV, we use the perturbative spectator model.
The scalar can decay to gluons with the width
\begin{align}
    \Gamma(\phi \to gg) = \frac{9\alpha_s^2(m_\phi)m_\phi^3 |c^{\text{eff}}_{gg}|^2}{72 \Lambda^2 \pi^3} \,,
\end{align}
where $\alpha_s$ is evaluated at the appropriate scale, or to quark-antiquark pairs with the decay width
\begin{align}
    \Gamma(\phi \to \bar{q} q)
            &=  \frac{3\varepsilon^2m_\phi |c^\ell_{qq}|^2 \beta_q^3}{16\pi}\,. 
\end{align}
In the mass range of interest, the most relevant channels are $\bar{s}s$ and $\bar{c}c$. 
Note that while decays to flavor-violating final states such as $\bar{c}u$ and $\bar{s}d$ are possible, they are subleading compared to the flavor-preserving final states, given the smallness of the flavor off-diagonal couplings in both the up and down sectors.

%%%%%%%%%%%%%%%%%%%%%%%%%%%%%%%%%%%%%%%%%
\section{$\NPfield$-Higgs Mixing}
\label{app:higgs_mixing}
%%%%%%%%%%%%%%%%%%%%%%%%%%%%%%%%%%%%%%%%%
A mixing between $\NPfield$ and the Higgs field would in principle introduce additional interactions between $\NPfield$ and the SM.
The general Lagrangian below the electroweak breaking scale, given by   \Eqs{eq:Lff_after_EWSB}{eq:LVV_after_EWSB},
becomes, in the presence of mixing,
\begin{align}
{\cal L}^\theta_{\NPfield\bar{f}f} &= \frac{\NPfield^{\text{\tiny phys}}}{\sqrt{2}} \left[ (\varepsilon c_{ij}^u+\theta \hat{Y}^u_{ij}) \bar{u}_i P_R u_j 
\! + \!  (\varepsilon c_{ij}^d+\theta \hat{Y}^d_{ij})   \bar{d}_i P_R d_j
\! + \!     (\varepsilon c_{ij}^\ell+\theta \hat{Y}^\ell_{ij}) \bar{\ell}_i P_R \ell_j
\! + \! \text{h.c.} \right] , \label{eq:Lff_after_EWSB_w_mixing}\\
 \! \! \!  {\cal L}^\theta_{\NPfield VV} &=  \frac{c_{gg}}{\Lambda}\frac{\alpha_s}{4\pi} \NPfield G^{\mu\nu}  G_{\mu\nu} +
    \frac{c_{\gamma\gamma}}{\Lambda}\frac{\alpha}{4\pi} \NPfield F^{\mu\nu}  F_{\mu\nu} +2\theta \frac{m_W^2}{v}W^+ W^-+... \label{eq:LVV_after_EWSB_w_mixing}\,,
\end{align}
where the mixing angle $\theta$ is defined so that the mass eigenstates are
\begin{align}
    \NPfield^{\text{\tiny phys}} &= \cos \theta \,\NPfield+\sin \theta \,h\,,
    \\      h^{\text{\tiny phys}} &= \cos \theta \,h-\sin \theta \,\NPfield\,.
\end{align}
Recall that $\hat Y^f$ are the diagonal Yukawa matrices.
In the FN model, the diagonal terms in $c^f$ and $\hat{Y}^f$ scale similarly.
Therefore, the effects of mixings in these terms are automatically taken into account in our analysis.
The mixing gives the dominant contribution if $\theta/\varepsilon \gg $1, and is sub-dominant otherwise.\footnote{There are several potential sources for $\sin\theta$, including a renormalizable Higgs-$\NPfield$ potential, heavy states at $\Lambda$,
and SM loops. The latter two give subleading effects  suppressed by additional powers of $v/\Lambda$.}

It also instructive to examine the finite off-diagonal couplings that are
generated at one loop due to the mixing~\cite{Boiarska:2019jym}:
\begin{align}
  \Delta  c^{\theta,d}  =   \frac{\theta}{\sqrt{2}}\frac{3}{32\pi^2} V^\dagger (\hat{Y}_u)^2 V \hat{Y}_d \,,
  \\
\Delta  c^{\theta,u}  =   \frac{\theta}{\sqrt{2}}\frac{3}{32\pi^2} V (\hat{Y}_d)^2 V^\dagger \hat{Y}_u \,.
\end{align}
The loop-induced mixing contribution to the $c \to u$ transition in the up sector is sub-dominant for
\begin{align}
   \theta/ \varepsilon \ll \frac{32 \pi^2 \text{Max}[c^u_{12},c^u_{21}]}{3  V_{ub} y_b^2 V^*_{cb}y_c } \sim 10^8 \frac{\text{Max}[c^u_{12},c^u_{21}]}{y_b^2}\,, 
\end{align}
where in the FN model $\text{Max}[c^u_{12},c^u_{21}] \sim \lambda^4 \sim y_b^2$.
This translates to an upper bound on the UV scale $\Lambda$,
\begin{align}
    \Lambda \ll  10^{14}\,\text{GeV}\,\left(\frac{10^{-4}}{\theta}\right)\frac{\text{Max}[c^u_{12},c^u_{21}]}{y_b^2}\,,
\end{align}
where we also plugged in a typical value of $\theta$ considered in the literature. 
This requirement 
is relaxed for $\mathcal{O}(1)$ couplings, in which case $\Lambda \ll  10^{17}(10^{-4}/ \theta)\,\text{GeV}$ is allowed.
Conversely, in the dark Higgs model, neglecting the off-diagonal contributions from dimension-5 operators to the $c \to u$ transition requires an exceptionally high UV scale.

Turning to the down sector, let us first consider the FNU models.
 The off-diagonal terms in the down sector are dominated by the dimension-5 operators if $\theta/\epsilon \ll \log (\Lambda^2/\mu_{\text{IR}}^2)$.

In the FN models, the mixing contribution to the $b \to s$ transition
can be neglected for
\begin{align}
   \theta/ \varepsilon \ll \frac{32  \pi ^2 }{3  V^*_{ts} y_t^2   V_{tb}  y_b } \sim  10^4 \frac{\text{Max}[c^d_{23},c^u_{32}]}{y_b} \,, 
\end{align}
where in the FN model $\text{Max}[c^d_{23},c^u_{32}] \sim \lambda^2 \sim y_b$.
For the dimension-5 contribution to dominate, the
upper bound on $\Lambda$ is
\begin{align}
     \Lambda \ll  10^{10}\,\text{GeV}\,\left(\frac{10^{-4}}{\theta}\right)\frac{\text{Max}[c^d_{23},c^u_{32}]}{y_b}\,.  
\end{align}
This requirement is relaxed for $\mathcal{O}(1)$ couplings, in which case $\Lambda \ll  10^{11}(10^{-4}/ \theta)\,\text{GeV}$ is 
allowed.
Since these transitions are less suppressed, neglecting the contribution from mixing
is consistent for lower values of $\Lambda$ compared to the up sector.
Conversely, in the dark Higgs model, neglecting the off-diagonal contributions to the $b\to s$ transition  from dimension-5 operators 
is consistent for lower UV scales compared to the $c\to u$ transition.

%%%%%%%%%%%%%%%%%%%%%%%%%%%%%%%%%%%%%%%%%%%%%%%%%%%%%%%%%%%%%%%%
\bibliographystyle{JHEP.bst}
\bibliography{FlavorAtFASER.bib}
%%%%%%%%%%%%%%%%%%%%%%%%%%%%%%%%%%%%%%%%%%%%%%%%%%%%%%%%%%%%%%%%

\end{document}